\documentclass{article}

\usepackage{arxiv}

\usepackage[utf8]{inputenc} 
\usepackage[T1]{fontenc}    
\usepackage{url}            
\usepackage{booktabs}       
\usepackage{amsfonts}       
\usepackage{nicefrac}       
\usepackage{microtype}      
\usepackage{lipsum}		
\usepackage{graphicx}
\usepackage[numbers]{natbib}
\usepackage{doi}
\usepackage{makecell, longtable}

\usepackage{amsmath}
\usepackage{amsthm}
\PassOptionsToPackage{table}{xcolor}
\usepackage{tikz,ifthen,fp,calc}
\usepackage{tikzpeople}
\usetikzlibrary{shapes}
\usepackage{fontawesome} 
\usepackage{pifont} 
\usepackage{wasysym} 
\usepackage[capitalize, noabbrev, nameinlink]{cleveref}
\usepackage{colortbl} 
\usepackage{pgfplots}
\pgfplotsset{compat=1.10}
\usepgfplotslibrary{fillbetween}
\usetikzlibrary{matrix}
\usetikzlibrary{patterns}
\usetikzlibrary{backgrounds}
\usepackage{subcaption}
\usepackage{multirow}
\usepackage{tcolorbox}

\usepackage{hyperref}
\hypersetup{colorlinks = true, citecolor = blue}

\pgfplotsset{scaled y ticks=false}

\newcommand{\PreserveBackslash}[1]{\let\temp=\\#1\let\\=\temp}
\newcolumntype{C}[1]{>{\PreserveBackslash\centering}p{#1}}
\newcolumntype{R}[1]{>{\PreserveBackslash\raggedleft}p{#1}}
\newcolumntype{L}[1]{>{\PreserveBackslash\raggedright}p{#1}}

\title{Network Analytics for Anti-Money Laundering -- A Systematic Literature Review and Experimental Evaluation}

\author{
	Bruno Deprez\thanks{Correspondening author: \href{mailto:bruno.deprez@kuleuven.be}{\texttt{bruno.deprez@kuleuven.be}}} \\
	KU Leuven\\
	University of Antwerp - imec
	\And
	Toon Vanderschueren\\
	KU Leuven\\
	University of Antwerp - imec
	\And
	Bart Baesens\\
	KU Leuven \\
	University of Southampton 
	\And
	Tim Verdonck\\
	University of Antwerp - imec\\
	KU Leuven 
	\And
	Wouter Verbeke\\
	KU Leuven
}

\begin{document}

\maketitle

\begin{abstract}
Money laundering presents a pervasive challenge, burdening society by financing illegal activities. 
The use of network information is increasingly being explored to effectively combat money laundering, given it involves connected parties. 
This led to a surge in research on network analytics for anti-money laundering~(AML). 
The literature is, however, fragmented and a comprehensive overview of existing work is missing.
This results in limited understanding of the methods to apply and their comparative detection power.
This paper presents an extensive and unique literature review, based on 97 papers from \textit{Web of Science} and \textit{Scopus}, resulting in a taxonomy following a recently proposed fraud analytics framework. 
We conclude that most research relies on expert-based rules and manual features, while deep learning methods have been gaining traction. 
This paper also presents a comprehensive framework to evaluate and compare the performance of prominent methods in a standardized setup. 
We compare manual feature engineering, random walk-based, and deep learning methods on two publicly available data sets. 
We conclude that (1) network analytics increases the predictive power, but caution is needed when applying GNNs in the face of class imbalance and network topology, and that (2) care should be taken with synthetic data as this can give overly optimistic results. 
The open-source implementation facilitates researchers and practitioners to extend this work on proprietary data, promoting a standardized approach for the analysis and evaluation of network analytics for AML. 
\end{abstract}

\keywords{Fraud Analytics \and Anti-Money Laundering \and Network Analytics \and Literature Review}

\section{Introduction}
Money laundering is the process of concealing illegally obtained funds by passing them through a complex sequence of transactions, so the money can be used to fund further criminal activities, e.g., drug trafficking, and terrorism~\citep{levi2006money}. 
The \citet{UNODC} estimates that around $2\%$ to $5\%$ of global GDP is laundered worldwide, amounting yearly to USD~2~trillion.

In contrast to credit card fraud, money laundering occurs over a longer duration and requires analysis of multiple transactions for detection.
It is actively concealed, since sustained laundering is needed to process continuous money streams. 
Credit card fraud, on the other hand, aims to get as much money out as fast as possible. 
Hence, the specific characteristics of money laundering merit dedicated research attention. 

Since this money needs to enter the financial system at some point~\citep{levi2006money}, legislators have put stringent rules on financial institutions~\citep{10.1214/ss/1042727940} for reporting suspicious transactions. Despite record fines and mounting pressure by regulators, money laundering represents a growing problem~\citep{williams2021fines}. 

Money laundering typically involves three stages~\citep{UNODC, levi2006money}: placement, where illicit funds enter the financial system; layering, where transactions obscure their origin; and integration, where the funds re-enter the economy through seemingly legitimate means, completing the laundering process.

Anti-money laundering practices require constant monitoring of clients and transactions.
These practices are, however, largely ineffective~\citep{economist2021losing}, with Europol estimating only 1.1\% of criminal profits in the EU being confiscated~\citep{europol2016does}.

Effective money laundering involves multiple parties, with individual transactions appearing normal~\citep{10.1214/ss/1042727940}. 
Therefore, network analytics is essential~\citep{10.1214/ss/1042727940, baesens2015fraud}. 
This need was already mentioned three decades ago by~\citet{senator1995financial}, where the authors used visualisations to support expert investing. 

The increasing attention of the past years has resulted in a growing body of literature on network analytics for anti-money laundering, but simultaneously resulted in a lack of comprehensive overviews.
Therefore, as a first step, this paper aims to supplement the literature by providing a comprehensive literature review of network analytics~(NA) for anti-money laundering, covering 97 papers analysed according to multiple dimensions.

Due to the fragmented nature of the literature there is limited insight into which methods perform best. 
This fragmentation and lack of comparison is apparent in the fact that many of the introduced methods are not tested against any baselines.
Furthermore, researchers are slow to adopt the latest network analytics methods and graph neural network in anti-money laundering. 
Therefore, the second goal of this work is to present a structured experimental set-up to evaluate the most prominent methods. 

In summary, our main contributions are four-fold: 
\begin{itemize}
    \item We provide an extensive and critical literature review on network analytics for anti-money laundering, to gain comprehensive and deep insights, and provide directions for future research.
    \item We benchmark a range of state-of-the-art network analytics methods for anti-money laundering, comparing the performance on two open-source data sets.
    \item We provide insights into the specific challenges of network methods based on the topology of the network and the data generating process.
    \item We implement the methods in a uniform manner and facilitate replication of the presented results by providing public access to the code at \url{https://github.com/B-Deprez/AML_Network}, aiming to promote a standardized approach towards the analysis and adoption of NA for AML and to encourage further research.
\end{itemize}

This paper is structured as follows. 
We present the methodology for the literature review in Section~\ref{sec:litrev}, with Section~\ref{sec:discussion} presenting the results and analysis. 
Section~\ref{sec:benchmark} describes the the empirical evaluation, with the results presented in Section~\ref{sec:res&disc}. 
Overall conclusions and directions for future work are presented in Section~\ref{sec:conclusion}.

\section{Literature Review}
\label{sec:litrev}
The first part of this work covers the current literature. 
A summary is given according to multiple categories. 
Additional analyses are provided for the top-cited papers. 
The aim is to provide answers to the following research questions:
\begin{itemize}
    \item What are the most prominent learning methods in the literature and how have these evolved over time? 
    \item How are methods evaluated, using what data sets and evaluation metrics? 
    \item What is the setting, objectives and challenges discussed in the top-cited papers? 
\end{itemize}

\subsection{Methodology}
\label{subsec:method}
We conducted the literature search using the queries \textit{“graph analy*” AND “money launder*”} and \textit{“network analy*” AND “money launder*”} on WoS and Scopus, chosen for their high completeness and research credibility~\citep{mongeon2016journal, bockel2023fraud}. 
The search is limited to English-language papers published before 2023, excluded recent publications to ensure adequate time for community evaluation.

The selection process included a title and abstract scan, followed by a full-text review. 
Papers were excluded if they focused on legal aspects or lacked substantial application of network analytics to anti-money laundering (AML). 
The final set of papers is classified according to seven criteria as summarised in Table~\ref{tab:classification}: publication data, method type, modelling approach, evaluation metrics, main objectives, data type, and network characteristics.

Our review provides both a broad analysis of trends across all selected papers and an in-depth focus on top-cited works (top $10\%$ annually and overall, per Google Scholar). 
This dual approach highlights general trends and the state-of-the-art (SOTA) in the field. 
Finally, we examine relevant review papers to contextualize our findings and demonstrate how our work extends existing literature.

\begin{table}[h]
\centering
\caption{The (sub-)categories with explanation for the classification of the literature.} 
\label{tab:classification}
    {\begin{tabular}{llL{9cm}}
    \toprule
         \textbf{Category} & \textbf{Sub-Category} & \textbf{Definition}  \\ \midrule
         \multirow{5}{*}{\textit{Publication Data}}& Title & The full title of the paper \\
          & Journal/Conference & Where the paper is published \\
          & Year & The year of publication \\
          & Review Paper & Indicating if it is a review paper \\
          & Bitcoin/Crypto & Indicating if it deals with crypto currency \\ \midrule
          \multirow{5}{*}{\textit{Method Type}} & Supervised & Application of supervised method \\
          & Unsupervised & Application of unsupervised method \\
          & Semi-Supervised & Application of semi-supervised method \\
          & Mixed & Methods presented across different learning methods\\
          & Visualisation & Application of visualisation method \\ \midrule
          \multirow{12}{*}{\textit{Modelling Method}}
          & Rule-based & Application of specific rules or cut-offs \\
          & Manual features & Network features having a exact definition \\
          & Walk-based & Network analysis based on (random) walks \\
          & Shallow representation & Embeddings typically through matrix factorization or random walks, without employing deep neural networks \\
          & Deep representation & Embeddings based on deep neural networks, e.g., GNNs \\
          & Correlation-based & Analysis of correlation of network features with target \\
          & Logistic regression & Application of logistic regression \\
          & Tree-based & Application of tree-based methods \\
          & SVM-based & Application of support vector machines \\
          & Neural networks & Learning using neural networks \\
          & Anomaly detection & Application of anomaly or outlier detection \\
          & Clustering & Features based on community detection or clustering of features \\
          \midrule
          \multirow{9}{*}{\textit{Evaluation Metric}} & Accuracy & Accuracy \\
          & Precision & The precision \\
          & Recall & The recall \\
          & F1 & The F1 or micro-F1 score \\
          & TPR & The true positive rate \\
          & FPR & The false positive rate \\
          & AUC-ROC & The area under the ROC curve \\
          & AUC-PR & The area under the precision-recall curve \\
          & Time & The execution time \\ \midrule
          \multirow{4}{*}{\textit{Objective}} & Client classification & Detection of suspicious entities \\
          & Transaction classification & Detection of suspicious transactions \\
          & Community detection & Detection of suspicious groups of clients \\
          & Flow/Chain detection & Detection of suspicious combination of transactions \\ \midrule
          \multirow{3}{*}{\textit{Data}} & Proprietary & Usage of confidential data \\
          & Open-source & Usage of freely available data \\
          & Synthetic & Usage of synthetically generated data \\ \midrule
          \multirow{2}{*}{\textit{Network}} & Multiple & Usage of multiple networks \\
          & Inverse/undirected & Additional use of network with directions removed or reverted \\
          \bottomrule
    \end{tabular}}
     {}
\end{table}

\begin{figure}
\centering
{\begin{tikzpicture}
	\node[] at (-0.1,-0.6) {Web of Science};
	\node[] at (2.5,-0.6) {Scopus};
	\draw [fill=lightgray!60,draw=none] (-1.4,0) rectangle (3.9,0.7) node[pos=.5] {Systematic Review};
	\draw [fill=lightgray!60,draw=none] (4.1,0) rectangle (7,0.7) node[pos=.5] {NA Methods};
	\draw [align = center] (-1.2,-2) rectangle (1.2,-1) node[pos=.5] {Query Search \\ (141)};
	\draw [align = center] (1.3,-2) rectangle (3.7,-1) node[pos=.5] {Query Search \\ (198)};
	\draw [align = center] (-1.2,-4) rectangle (1.2,-2.5) node[pos=.5] {Title \& \\ Abstract Scan \\ (84)};
	\draw [align = center] (1.3,-4) rectangle (3.7,-2.5) node[pos=.5] {Title \& \\ Abstract Scan \\ (103)};
	\draw [align = center] (-0.5,-5.5) rectangle (3,-4.5) node[pos=.5] {Full Text Scan \\ (97)};
	\draw[->] (-0.1,-2) -- (-0.1,-2.49);
	\draw[->] (2.5,-2) -- (2.5,-2.49);
	\draw[->] (-0.1,-4) -- (-0.1,-4.49);
	\draw[->] (2.5,-4) -- (2.5,-4.49);
	\draw [align = center] (-0.5,-7) rectangle (3,-6) node[pos=.5] {Top Cited \\ (20)};
	\draw[->] (1.2,-5.5) -- (1.2,-5.99);
	\draw [align = center] (4.2,-2) rectangle (6.8,-1) node[pos=.5] {Wider Search};
	\draw [align = center] (4.2,-7) rectangle (6.8,-6) node[pos=.5] {State-of-the-Art \\ (Section~\ref{subsec:model})};
	\draw[->] (5.5,-2) -- (5.5,-5.99);
	\draw [align = center]  (-0.5,-8.5) rectangle (6.8,-7.5) node[pos=.5] {Experimental Evaluation  (Section~\ref{sec:benchmark})};
	\draw[->] (5.5,-7) -- (5.5,-7.49);
	\draw[->] (1.2,-7) -- (1.2,-7.49);
\end{tikzpicture} }
\caption{The search and filtering method with the number of remaining papers between brackets.}
\label{fig:flow}
{}
\end{figure}

\subsection{Results}
\label{subsec:results}
Figure~\ref{fig:flow} summarizes the initial search results. 
After filtering and removing duplicates, 97 papers remained, of which 11 were review papers. 
Classification based on Table~\ref{tab:classification} is detailed in Table~\ref{tab:literature-1} and Table~\ref{tab:literature-2}, with review papers excluded from category-specific analysis due to their broader scope.

\begin{scriptsize}
\renewcommand*{\arraystretch}{1.05} 
\setlength{\tabcolsep}{3pt} 
\centering
\begin{longtable}{C{5cm}| C{1cm}C{0.31cm}C{0.31cm} |  C{0.30cm}C{0.30cm}C{0.30cm}C{0.30cm}C{0.30cm}C{0.30cm}C{0.30cm}C{0.30cm}C{0.30cm} | C{0.30cm}C{0.30cm}C{0.30cm}C{0.30cm} | C{0.30cm}C{0.30cm}C{0.30cm} | C{0.30cm}C{0.30cm}}
\caption{\textbf{Literature classification.} Classification of the papers according to different categories: publication data, learning method, evaluation metric, objective, and data.}
\label{tab:literature-1} \\
\toprule
&\multicolumn{3}{ c|}{\bf Publication Data} &\multicolumn{9}{ c|}{\bf Evaluation Metric}&\multicolumn{4}{ c|}{\bf Objective}&\multicolumn{3}{c|}{\bf Data} &\multicolumn{2}{ c}{\bf Network} \\
\rotatebox[origin=l]{90}{\textit{Reference}} & \rotatebox[origin=l]{90}{\textit{Year}} & \rotatebox[origin=l]{90}{\textit{Review Paper}} & \rotatebox[origin=l]{90}{\textit{Bitcoin/Crypto}} & \rotatebox[origin=l]{90}{\textit{Accuracy}} & \rotatebox[origin=l]{90}{\textit{Precision}} & \rotatebox[origin=l]{90}{\textit{Recall}} & \rotatebox[origin=l]{90}{\textit{F1}} & \rotatebox[origin=l]{90}{\textit{TPR}} & \rotatebox[origin=l]{90}{\textit{FPR}} & \rotatebox[origin=l]{90}{\textit{AUC-ROC}} & \rotatebox[origin=l]{90}{\textit{AUC-PR}} & \rotatebox[origin=l]{90}{\textit{Time}} & \rotatebox[origin=l]{90}{\textit{Client Classification}} & \rotatebox[origin=l]{90}{\textit{Transaction Classification} } & \rotatebox[origin=l]{90}{\textit{Community Detection}} & \rotatebox[origin=l]{90}{\textit{Flow/Chain Detection}} & \rotatebox[origin=l]{90}{\textit{Proprietary}} & \rotatebox[origin=l]{90}{\textit{Open-Source}} & \rotatebox[origin=l]{90}{\textit{Synthetic}}
& \rotatebox[origin=l]{90}{\textit{Inverse/undirected network}} & \rotatebox[origin=l]{90}{\textit{Multiple networks}}\\
\midrule
\endfirsthead
\multicolumn{22}{l}{\normalsize Table \thetable\ (continued)} \\
\toprule
&\multicolumn{3}{ c|}{\bf Publication Data} &\multicolumn{9}{ c|}{\bf Evaluation Metric}&\multicolumn{4}{ c|}{\bf Objective}&\multicolumn{3}{c|}{\bf Data} &\multicolumn{2}{ c}{\bf Network} \\
\rotatebox[origin=l]{90}{\textit{Reference}} & \rotatebox[origin=l]{90}{\textit{Year}} & \rotatebox[origin=l]{90}{\textit{Review Paper}} & \rotatebox[origin=l]{90}{\textit{Bitcoin/Crypto}} & \rotatebox[origin=l]{90}{\textit{Accuracy}} & \rotatebox[origin=l]{90}{\textit{Precision}} & \rotatebox[origin=l]{90}{\textit{Recall}} & \rotatebox[origin=l]{90}{\textit{F1}} & \rotatebox[origin=l]{90}{\textit{TPR}} & \rotatebox[origin=l]{90}{\textit{FPR}} & \rotatebox[origin=l]{90}{\textit{AUC-ROC}} & \rotatebox[origin=l]{90}{\textit{AUC-PR}} & \rotatebox[origin=l]{90}{\textit{Time}} & \rotatebox[origin=l]{90}{\textit{Client Classification}} & \rotatebox[origin=l]{90}{\textit{Transaction Classification} } & \rotatebox[origin=l]{90}{\textit{Community Detection}} & \rotatebox[origin=l]{90}{\textit{Flow/Chain Detection}} & \rotatebox[origin=l]{90}{\textit{Proprietary}} & \rotatebox[origin=l]{90}{\textit{Open-Source}} & \rotatebox[origin=l]{90}{\textit{Synthetic}}
& \rotatebox[origin=l]{90}{\textit{Inverse/undirected network}} & \rotatebox[origin=l]{90}{\textit{Multiple networks}}\\
\midrule
\endhead
\bottomrule
\multicolumn{22}{r}{\small\textit{Continued on next page}} \\
\endfoot
\endlastfoot
Total & & 11 & 26 & 9 & 15 & 18 & 17 & 4 & 5 & 10 & 1 & 11 & 37 & 27 & 13 & 16 & 32 & 40 & 11 & 5 & 18  \\
\midrule
\citet{DREZEWSKI201518} & 2015& \textcolor{gray!25}{\ding{55}} &\textcolor{gray!25}{\ding{55}} & \textcolor{gray!25}{\ding{55}} & \textcolor{gray!25}{\ding{55}} & \textcolor{gray!25}{\ding{55}} & \textcolor{gray!25}{\ding{55}} & \textcolor{gray!25}{\ding{55}} & \textcolor{gray!25}{\ding{55}} & \textcolor{gray!25}{\ding{55}} & \textcolor{gray!25}{\ding{55}} & \ding{51} & \ding{51} & \textcolor{gray!25}{\ding{55}} & \textcolor{gray!25}{\ding{55}} & \textcolor{gray!25}{\ding{55}} & \ding{51} & \textcolor{gray!25}{\ding{55}} & \textcolor{gray!25}{\ding{55}} & \textcolor{gray!25}{\ding{55}} & \textcolor{gray!25}{\ding{55}}\\
\rowcolor{gray!10}\citet{NGAI2011559} & 2011& \ding{51} &\textcolor{gray!25}{\ding{55}} &   &   &   &   &   &   &   &   &   &   &   &   &   &   &   &   &   &  \\
\citet{malm2013using} & 2013& \textcolor{gray!25}{\ding{55}} &\textcolor{gray!25}{\ding{55}} & \textcolor{gray!25}{\ding{55}} & \textcolor{gray!25}{\ding{55}} & \textcolor{gray!25}{\ding{55}} & \textcolor{gray!25}{\ding{55}} & \textcolor{gray!25}{\ding{55}} & \textcolor{gray!25}{\ding{55}} & \textcolor{gray!25}{\ding{55}} & \textcolor{gray!25}{\ding{55}} & \textcolor{gray!25}{\ding{55}} & \ding{51} & \textcolor{gray!25}{\ding{55}} & \textcolor{gray!25}{\ding{55}} & \textcolor{gray!25}{\ding{55}} & \textcolor{gray!25}{\ding{55}} & \ding{51} & \textcolor{gray!25}{\ding{55}} & \textcolor{gray!25}{\ding{55}} & \textcolor{gray!25}{\ding{55}}\\
\bottomrule
\end{longtable}
\end{scriptsize}


\begin{scriptsize}
\renewcommand*{\arraystretch}{1.05} 
\setlength{\tabcolsep}{3pt} 
\centering
\begin{longtable}{C{5cm}| C{1.2cm}C{0.8cm} |C{0.31cm}C{0.31cm}C{0.31cm}C{0.31cm}C{0.31cm} |C{0.31cm}C{0.31cm}C{0.31cm}C{0.31cm}C{0.31cm}C{0.31cm}C{0.31cm}C{0.31cm}C{0.31cm}C{0.31cm}C{0.31cm}C{0.31cm} }
\caption{\textbf{Literature classification.} Classification of the papers according to the different method building blocks.}
\label{tab:literature-2}\\
\toprule
&\multicolumn{2}{ c|}{\bf Publication Data} &\multicolumn{5}{ c|}{\textbf{Method Type}} &\multicolumn{12}{ c}{\textbf{Modelling Method}}\\
\rotatebox[origin=l]{90}{\textit{Reference}} & \rotatebox[origin=l]{90}{\textit{Year}} & \rotatebox[origin=l]{90}{\textit{Bitcoin/Crypto}} &  \rotatebox[origin=l]{90}{\textit{Supervised}}& \rotatebox[origin=l]{90}{\textit{Unsupervised}}& \rotatebox[origin=l]{90}{\textit{Semi-Supervised}}& \rotatebox[origin=l]{90}{\textit{Mixed}}& \rotatebox[origin=l]{90}{\textit{Visualisation}}&\rotatebox[origin=l]{90}{\textit{Rule-Based}} & \rotatebox[origin=l]{90}{\textit{Manual Features}} & \rotatebox[origin=l]{90}{\textit{Walk-based}} & \rotatebox[origin=l]{90}{\textit{Shallow Repr.}} & \rotatebox[origin=l]{90}{\textit{Deep Repr.}} & \rotatebox[origin=l]{90}{\textit{Correlation-Based}} & \rotatebox[origin=l]{90}{\textit{Tree-Based}} & \rotatebox[origin=l]{90}{\textit{SVM-Based}} & \rotatebox[origin=l]{90}{\textit{Neural Networks}} & \rotatebox[origin=l]{90}{\textit{Anomaly Detection}} & \rotatebox[origin=l]{90}{\textit{Clustering}} & \rotatebox[origin=l]{90}{\textit{Logistic Regression}}\\
\midrule
\endfirsthead
\multicolumn{20}{l}{\normalsize Table \thetable\ (continued)} \\
\toprule
&\multicolumn{2}{ c|}{\bf Publication Data} &\multicolumn{5}{ c|}{\textbf{Method Type}} &\multicolumn{12}{ c}{\textbf{Modelling Method}}\\
\rotatebox[origin=l]{90}{\textit{Reference}} & \rotatebox[origin=l]{90}{\textit{Year}} & \rotatebox[origin=l]{90}{\textit{Bitcoin/Crypto}} &  \rotatebox[origin=l]{90}{\textit{Supervised}}& \rotatebox[origin=l]{90}{\textit{Unsupervised}}& \rotatebox[origin=l]{90}{\textit{Semi-Supervised}}& \rotatebox[origin=l]{90}{\textit{Mixed}}& \rotatebox[origin=l]{90}{\textit{Visualisation}}&\rotatebox[origin=l]{90}{\textit{Rule-Based}} & \rotatebox[origin=l]{90}{\textit{Manual Features}} & \rotatebox[origin=l]{90}{\textit{Walk-based}} & \rotatebox[origin=l]{90}{\textit{Shallow Repr.}} & \rotatebox[origin=l]{90}{\textit{Deep Repr.}} & \rotatebox[origin=l]{90}{\textit{Correlation-Based}} & \rotatebox[origin=l]{90}{\textit{Tree-Based}} & \rotatebox[origin=l]{90}{\textit{SVM-Based}} & \rotatebox[origin=l]{90}{\textit{Neural Networks}} & \rotatebox[origin=l]{90}{\textit{Anomaly Detection}} & \rotatebox[origin=l]{90}{\textit{Clustering}} & \rotatebox[origin=l]{90}{\textit{Logistic Regression}}\\
\midrule
\endhead
\bottomrule
\multicolumn{20}{r}{\small\textit{Continued on next page}} \\
\endfoot
\endlastfoot
Total & & 26 & 26 & 40 & 6 & 3 & 20 & 20 & 45 & 15 & 6 & 12 & 3 & 10 & 4 & 6 & 10 & 24 & 6 \\ 
\midrule
\citet{DREZEWSKI201518} & 2015& \textcolor{gray!25}{\ding{55}} & \textcolor{gray!25}{\ding{55}} & \ding{51} & \textcolor{gray!25}{\ding{55}} & \textcolor{gray!25}{\ding{55}} & \textcolor{gray!25}{\ding{55}} & \ding{51} & \ding{51} & \textcolor{gray!25}{\ding{55}} & \textcolor{gray!25}{\ding{55}} & \textcolor{gray!25}{\ding{55}} & \textcolor{gray!25}{\ding{55}} & \textcolor{gray!25}{\ding{55}} & \textcolor{gray!25}{\ding{55}} & \textcolor{gray!25}{\ding{55}} & \textcolor{gray!25}{\ding{55}} & \textcolor{gray!25}{\ding{55}} & \textcolor{gray!25}{\ding{55}}\\
\rowcolor{gray!10}\citet{malm2013using} & 2013& \textcolor{gray!25}{\ding{55}} & \textcolor{gray!25}{\ding{55}} & \ding{51} & \textcolor{gray!25}{\ding{55}} & \textcolor{gray!25}{\ding{55}} & \ding{51} & \textcolor{gray!25}{\ding{55}} & \ding{51} & \textcolor{gray!25}{\ding{55}} & \textcolor{gray!25}{\ding{55}} & \textcolor{gray!25}{\ding{55}} & \textcolor{gray!25}{\ding{55}} & \textcolor{gray!25}{\ding{55}} & \textcolor{gray!25}{\ding{55}} & \textcolor{gray!25}{\ding{55}} & \textcolor{gray!25}{\ding{55}} & \textcolor{gray!25}{\ding{55}} & \textcolor{gray!25}{\ding{55}}\\
\citet{FRONZETTICOLLADON201749} & 2017& \textcolor{gray!25}{\ding{55}} & \ding{51} & \textcolor{gray!25}{\ding{55}} & \textcolor{gray!25}{\ding{55}} & \textcolor{gray!25}{\ding{55}} & \textcolor{gray!25}{\ding{55}} & \textcolor{gray!25}{\ding{55}} & \ding{51} & \textcolor{gray!25}{\ding{55}} & \textcolor{gray!25}{\ding{55}} & \textcolor{gray!25}{\ding{55}} & \ding{51} & \textcolor{gray!25}{\ding{55}} & \textcolor{gray!25}{\ding{55}} & \textcolor{gray!25}{\ding{55}} & \textcolor{gray!25}{\ding{55}} & \textcolor{gray!25}{\ding{55}} & \textcolor{gray!25}{\ding{55}}\\
\bottomrule
\end{longtable}
\end{scriptsize}

\begin{figure}
    \centering
    {\includegraphics[width = 0.6\textwidth]{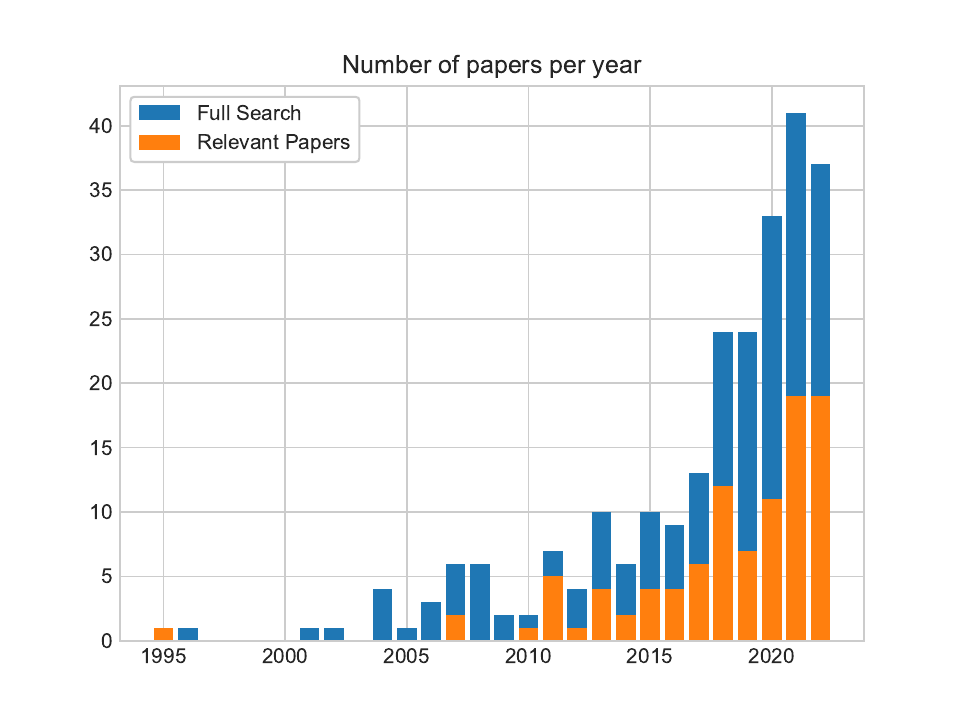}}
    \caption{The number of papers incorporated in the full query and after filtering per year.} 
    \label{fig:yearlyNumber}
    {}
\end{figure}

Figure~\ref{fig:yearlyNumber} illustrates the increasing trend in AML-related research. 
Table~\ref{tab:journals} lists journals and conferences with more than one relevant publication, covering 28 papers; the remaining 69 papers were distributed across other venues with a single publication each. 
The journal with most publications is the Journal of Money Laundering Control, with seven published papers. 
On second place is the ASONAM conference with three papers on network analytics for AML.

Crypto-related research accounts for 26 papers ($26.5\%$ of the total), analysed separately in Appendix~A due to its distinct AML setting, data characteristics, and anonymity challenges. 
These differences are evident in the corresponding plots.

Appendix~B presents results for the 20 top-cited papers, summarized in Table~\ref{tab:topliterature}. 
The category distribution aligns with the overall literature, with $25\%$ focusing on cryptocurrency, closely matching the full-scope percentage.

\begin{table}[t]
    \centering
    \caption{Journals/conferences that have two or more publications in scope of this review.}
    \label{tab:journals}
    {
    \begin{tabular}{l|c}
    \toprule
\bf Journal/Conference Title & \bf Number\\ 
\midrule
    Journal of Money Laundering Control & 7\\
\rowcolor{gray!10}International Conference on Advances in Social Network Analysis and Mining, ASONAM & 3\\
CEUR Workshop Proceedings & 2\\
\rowcolor{gray!10}IEEE International Conference on Data Mining Workshops & 2\\
IEEE International Conference on Big Data (Big Data) & 2\\
\rowcolor{gray!10}Journal of Physics: Conference Series & 2\\
Federated Conference on Computer Science and Information Systems (FedCSIS) & 2\\
\rowcolor{gray!10}EPJ Data Science & 2\\
Information Sciences & 2\\
\rowcolor{gray!10}IEEE Access & 2\\
International Conference on Machine Learning Technologies & 2\\
    \bottomrule
    \end{tabular}}
{}
\end{table}

\section{Discussion on the Literature}
\label{sec:discussion}
\subsection{Review Papers}
\label{subsec:revpapers}
We identified 11 review papers, which we discuss first to illustrate how our work complements and extends the existing literature. 

\citet{NGAI2011559,kurshan2020graph} and \citet{lokanan2022financial} cover general financial fraud, with less emphasis on anti-money laundering or graph learning. 
\citet{NGAI2011559} construct a classification framework, but include only one paper on money laundering, which is the only paper on network analytics. 
\citet{kurshan2020graph} give a high-level overview of the general difficulties of machine learning solutions for fighting financial crime. 
They noted that network-based methods have a lot of potential to process many transactions handled by financial institutions. 
The paper is closer to a discussion paper than to a systematic literature review. 
The scope of the review by~\citet{lokanan2022financial} is limited, covering only visualisation methods for outlier detection. 

Other review papers deal specifically with AML~\citep{kute2021deep, semenov2017survey, gao2007framework}. 
\citet{kute2021deep} give an overview of the methods applied in the literature, but their main topic is interpretability. 
\citet{semenov2017survey} discuss the different methods applied to AML. 
It touches upon different network methods, but it is not a systematic literature review. 
The framework by~\citet{gao2007framework} again deals with a wider array of methods, with the authors concluding that network analytics is the most promising. 
However, the number of papers on network analytics for AML included is limited. 

\citet{gerbrands2022effect} look at the effects of money laundering policies, by measuring the change in network features when a major AML law is introduced. 
It analyses cluster size and the degree, closeness and betweenness centrality. 
Hence, the scope is different, since we focus on enhancing anti-money laundering using network features. 

A significant number of review papers deal with crypto currency~\citep{irwin2018illicit, day2021artificial, alarab2020comparative, han2021blockchain}.
\citet{irwin2018illicit} describe challenges in detecting money laundering on the blockchain. 
This view results in a narrow scope with a total of 45 unique references. 
These different challenges are partially addressed by~\citet{alarab2020comparative} and~\citet{han2021blockchain}. 
The experiment by~\citet{alarab2020comparative} compares different supervised machine learning methods, with the statistics of the ego-network of the different nodes in the Bitcoin transaction network~\citep{Elliptic, weber2019anti} as features. 

Other papers cover a specific money laundering method and propose tailor-made solutions. 
\citet{han2021blockchain} discuss the application of anomaly detection, based on crime-specific transaction patterns. 
The paper also covers Ponzi schemes and blackmail campaigns. 
Their scope is more limited, with a total of 33 references. 
Similarly, \citet{day2021artificial} gives an overview of different characteristics of financial crime, and uses these to construct knowledge graphs for building legal cases. 

We extend on prior work by conducting the first systematic review of network analytics methods for anti-money laundering, including both supervised and unsupervised methods. 
Our scope spans traditional centrality measures to advanced graph neural networks, and addresses money laundering in both fiat and crypto currency.

\subsection{Full Scope of Papers}
\label{subsec:fullscope}
This section summarises the 83 non-review papers. 
Although network analytics for anti-money laundering has been researched for a long time, Figure~\ref{fig:yearlyNumber} illustrates this increasing research interest only after 2010. 
The average increase in output is $61\%$ year-on-year for the papers in scope. 

\subsubsection{Methods.} The methods in the literature are supervised, unsupervised, semi-supervised and visualisations. 
The most popular method is unsupervised learning, followed by supervised learning and visualisations, as shown in Table~\ref{tab:literature-2}. 
The popularity of unsupervised learning may be due to money laundering being very uncommon, resulting in highly skewed data sets with very few labelled observations~\citep{akoglu2015graph}. 
Due to limited resources to investigate the massive amounts of transactions, most transactions are left unlabelled~\citep{van2023catchm,DBLP:journals/corr/abs-1009-6119}. 
In addition, criminals try to evade detection by continuously changing tactics~\citep{baesens2015fraud,DBLP:journals/corr/abs-1009-6119}, making historically patterns less relevant for future predictions~\citep{10.1214/ss/1042727940}. 
The performance of (supervised) models trained on historical data can therefore decrease over time~\citep{bockel2023fraud}. 

Table~\ref{tab:literature-2} indicates that half of the papers rely on manual feature engineering. These manual features include classic network centrality measures (degree, betweenness etc.) and summary statistics of the transactions in the nodes' ego-network.
The second most popular method is clustering, which is used broadly in two ways. 
First, clustering for visualisations mitigates visual cluttering and allows the the selection detail shown. 
Second, clusters are used for feature engineering. 
Next to manual feature engineering and clustering, rule-based methods are still frequently used. 
This illustrates that the current literature heavily relies on expert knowledge. 
Hence, there is an opportunity to further develop data-driven machine learning methods in anti-money laundering.

\begin{figure}[t]
			\centering
			{\includegraphics[width = \textwidth]{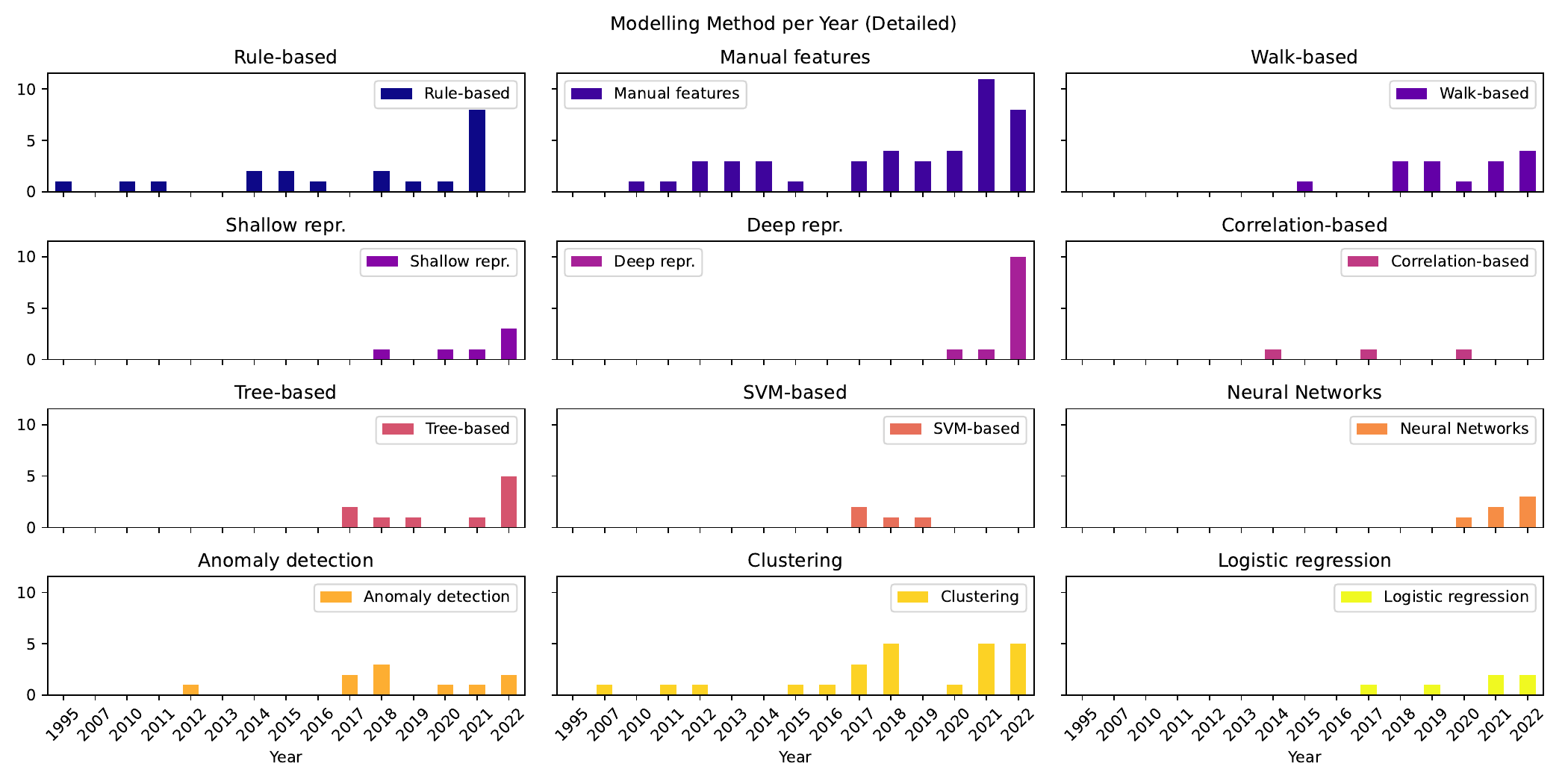}}
			\caption{The evolution of the model building blocks over the years. Since there are many modelling methods, the graphs are split.}
            \label{fig:disttechyear_detail}
   {}
		\end{figure}

The evolution of methods used over time, as presented in Figure~\ref{fig:disttechyear_detail}, shows that the earliest research relied almost exclusively on rule-based systems and manual feature engineering. 
These remain popular approaches to this day. 
From 2018 onward, we see the application of deep learning methods based on shallow representations and graph neural networks. 
Especially the use of GNNs has seen a spike in 2022.

The downstream methods used for classification have also evolved. 
Popular methods are tree-based, clustering, anomaly detection and logistic regression. 
SVM-based methods were popular until 2019, while neural networks were only used in AML as of 2020. 
These are often recurrent neural networks to incorporate the temporal aspect~\citep{xia2021novel, zhang2021dyngraphtrans,mohan2022improving,li2022transactional}.

Key to these methods is how the authors define the network(s) used. 
The construction of multiple networks, e.g., same entities but using different relations, to capture richer data is present in 18 papers. 
Of these papers, five used the undirected network or the one with direction reversed in tandem with the original network. 
It allows for information propagate from, e.g., the transaction's receiver back to the sender.

\subsubsection{Objectives.}
Table~\ref{tab:literature-1} shows that the AML method's objective is most often detecting suspicious clients, followed by detecting suspicious transactions, money flows and communities. 
Client classification combines client characteristics with payments made over a longer period. 
Since money laundering is done over a longer period, with the aim to make the individual transaction appear normal, models trained on a series of transactions, leveraging an individual's behaviour, can obtain higher performance. 

Figure~A3 shows that methods for crypto-currency are most often introduced for suspicious transaction detection. 
Due to the pseudo-anonymity, it is much harder to know which wallets belong to what person~\citep{zhou2022behavior, XUESHUO2021107507}, making it less feasible to build profiles. 
Therefore, researchers in AML have also looked at de-anonymisation of the wallets~\citep{zhou2022behavior,XUESHUO2021107507,gong2022analyzing}. 
They use network features to cluster wallets together to see if they belong to the same person or organization. 

In recent years, more studies have been done on flow/chain detection, tracking payments over multiple steps/people. 
Although this requires more computing power, it more closely resembles reality~\citep{UNODC, levi2006money}. 
Hence, these methods are becoming more widely used. 

\subsubsection{Evaluation.}
\begin{figure}[t]
			\centering
			{\includegraphics[width = \textwidth]{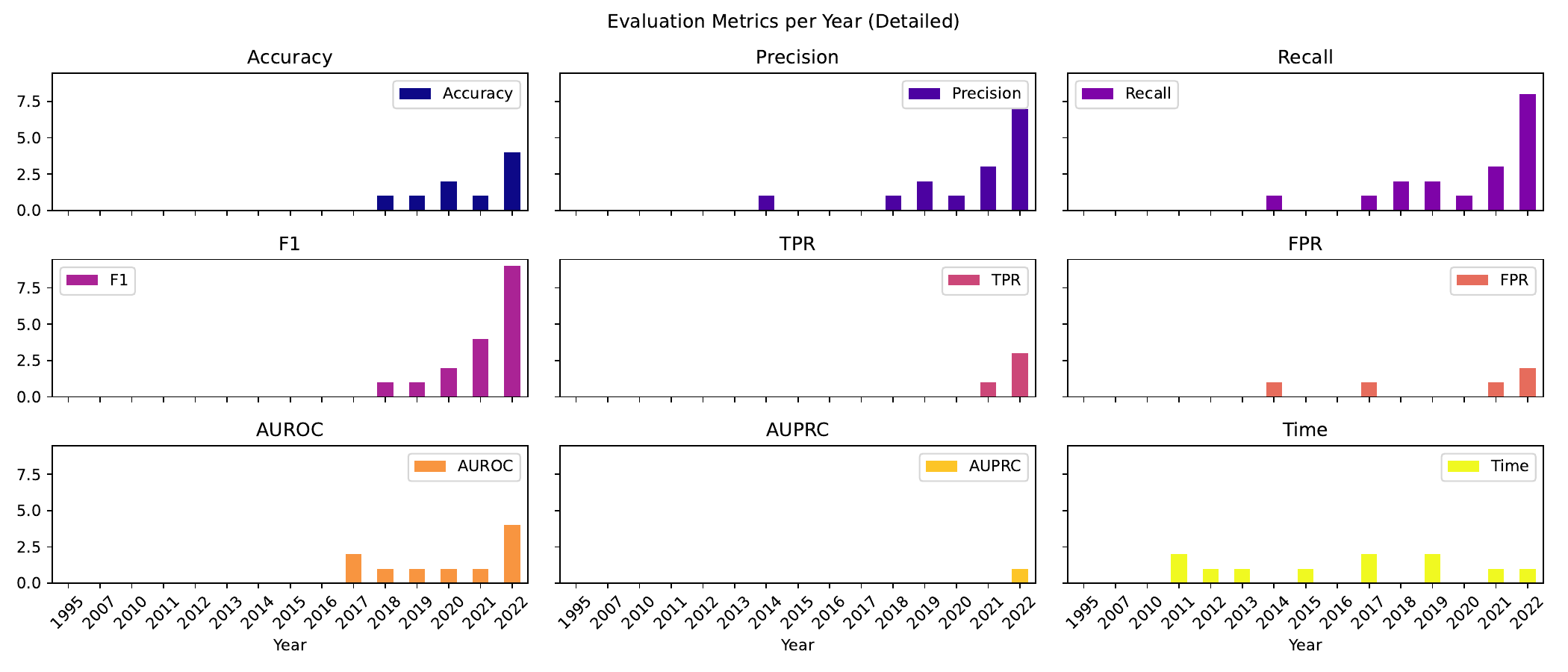}}
			\caption{The evolution of the evaluation metrics over the years. Since there are many metrics, the graphs are split.}
            \label{fig:distmetyear}
   {}
		\end{figure}

Table~\ref{tab:literature-1} shows that research mostly relies on the recall, $F1$-score, precision and time. 
Figure~\ref{fig:distmetyear} illustrates that training/prediction time was a popular method in earlier studies, but relatively few papers use time in more recent work, which might be a consequence of increased computing power. 
Additionally, the popularity of the precision, recall and $F1$-score is a recent and increasing trend in the literature. 

We remark two surprising results. First, only one paper used the area under the precision-recall curve~(AUC-PR). 
Due to the high label imbalanced, this metric is shown to be more appropriate than the area under the ROC curve~(AUC-ROC)~\citep{davis2006relationship, ozenne2015precision}. 
Therefore, we would expect it to be more widely used. 
Second, some papers solely reported accuracy~\citep{DIDIMO2019406, xiao2018visual, gong2022analyzing,xiong2020identification}. 
Given the high class imbalance, accuracy fails in providing an appropriate performance assessment.

\subsubsection{Data.}
Table~\ref{tab:literature-1} shows that most studies use open-source data, with proprietary data used in almost $50\%$ of the papers. 
Only a minority of papers report their findings using synthetic data. 
There is, however, a major difference between crypto- and other research. 
Figure~A1 shows that crypto-research is almost exclusively done on open-source data. 
This results from the blockchain's freely accessible distributed ledger~\citep{weber2019anti}. 
On the other hand, \textit{classic} transaction data cannot be shared, due to privacy reasons, resulting in half the data sets being proprietary.

Following this last remark, one would expect to have (almost) no research done on open-source data sets for fiat currencies.
Figure~A1 shows that almost $30\%$ of non-crypto related papers use open-source data. 
These papers can be categorised in four groups:
\begin{itemize}
    \item Methods that are evaluated on non-financial network data, whether or not next to proprietary/synthetic money laundering data:~\citet{10.1145/3097983.3098015, ovelgonne2012covertness,micale2019fast,prado2017community}
    \item Methods evaluated on government data, assumed  or stated to be open-source:~\citet{didimo2011advanced, IMANPOUR2019105856,velasco2021decision, bahulkar2018integrative, malm2013using}
    \item Methods constructed to analyse leaked documents/scandals, e.g., the Panama papers:~\citet{joaristi2018inferring,joaristi2019detecting,MAGALINGAM20151,adriaens2019discovering,winiecki2020validating,helmy2016design,cheong2010event}
    \item Other:~\citet{nandhini2016assessment,bhalerao2019mapping,ccauglayan35money}
\end{itemize}
For the \textit{Other} category, we have the following. 
\citet{nandhini2016assessment} say that the data is publicly available, without giving any more information. 
\citet{bhalerao2019mapping} scraped their data from Hack~Forums. 
Finally, \citet{ccauglayan35money} use a Kaggle data set based on synthetic data. 
We categorise it as open-source, since it is made publicly available. 

Table~\ref{tab:data sets} summarises the open-source data sets. 
The most widely used are almost all crypto-currency data sets. 
Raw Bitcoin data is used by ten papers. 
This raw data is often enhanced using data from other platforms. 
One such platform is WalletExplorer, which is used by three papers. 
It is less relevant today, since the data has not been maintained since 2016. 
Another recent data set on Bitcoin is the Elliptic data sets~\citep{Elliptic,weber2019anti}. 
Although only published in 2019, it has already been used in six experiments. 
After Bitcoin, Ethereum seems to be the second most popular crytpo-currency in AML.

The most used data set for fiat currency is the Czech Financial Data set (CFD). 
This data set does not provide any labels. 
Papers using it introduce additional, synthetic money laundering patterns to test their methods~\citep{li2020flowscope,wang2022detection}. 
This is not ideal, as this hinders comparison across research and can introduce detection bias.

As mentioned above, AML methods are also being evaluated on leaked data like the ICIJ Offshore Leaks Database (including, e.g., the Panama-papers) and the Enron data set. 

\begin{table}[t]
    \centering
    \caption{The open-source data sets used in the different papers.}
    \label{tab:data sets}
    {
    \begin{tabular}{l|cC{8cm}}
    \toprule
\bf Data Set & \bf Number & \bf References\\ 
\midrule
Raw bitcon                            & 10 &  \citep{mcginn2016visualizing}, \citep{wu2021detecting}, \citep{phetsouvanh2018egret}, \citep{wu2021towards}, \citep{turner2020discerning}, \citep{ricci2020economic}, \citep{XUESHUO2021107507}, \citep{tharani2021graph}, \citep{rathore2022mixers}, \citep{yang2022illicit} \\
Elliptic Data set                       & 6 & \citep{alarab2020competence}, \citep{xia2021novel}, \citep{mohan2022improving}, \citep{sun2022game}, \citep{li2022blockchain}, \citep{li2022transactional} \\
Ethereum XBlock.pro                    & 4 & \citep{tharani2021graph}, \citep{zhou2022behavior}, \citep{jin2022heterogeneous}, \citep{jin2022detecting} \\
ICIJ Offshore Leaks Database           & 4 & \citep{joaristi2019detecting}, \citep{garcia2021ai}, \citep{joaristi2018inferring}, \citep{winiecki2020validating} \\
WalletExplorer                         & 3 & \citep{wu2021detecting}, \citep{turner2020discerning}, \citep{XUESHUO2021107507} \\
Czech Financial Data set (CFD)          & 2 & \citep{li2020flowscope}, \citep{wang2022detection} \\
Enron Data set                          & 2 & \citep{MAGALINGAM20151}, \citep{adriaens2019discovering} \\
RenRenDai                              & 1 & \citep{xiao2018visual} \\
Bloxy                                  & 1 & \citep{jin2022detecting} \\
iCOV                                   & 1 & \citep{gerbrands2022effect}\\
Secret Presidential Funds scandal      & 1 & \citep{cheong2010event} \\
Tencent QQ                             & 1 & \citep{zhou2017analyzing} \\
Caviar data set                        & 1 & \citep{bahulkar2018integrative} \\
BitcoinTalk                            & 1 & \citep{wu2021towards} \\
Ripple Network                         & 1 & \citep{camino2017finding} \\
2015 Ashley Madison extortion scam     & 1 & \citep{phetsouvanh2018egret} \\
FinCEN                                 & 1 & \citep{didimo2011advanced} \\
BlockCypher                            & 1 & \citep{rathore2022mixers} \\
    \bottomrule
    \end{tabular}}
{}
\end{table}

\subsection{Top Cited Papers}
\label{subsec:top}
A final step in our discussion is an in-depth analysis of the top-cited papers. 
These papers are well received by the scientific community and are likely of high quality. 
We select the $10\%$ most cited papers per year, extended with the $10\%$ most cited papers overall (see Section~\ref{subsec:method}). 
This results in 20 papers as listed in Table~\ref{tab:topliterature}. 
The analysis is done in three stages to gain deeper insights. 
First, we revisit the categories as presented in Table~\ref{tab:classification}. 
Second, we analyse the task-level objectives by discussing the data and methods present in these literature. 
Third, the papers are classified according to the processing steps and challenges, as introduced by~\citet{bockel2023fraud}.

\begin{table}[t]
\centering
\caption{Summary of the top-cited papers, with indication of whether it is a review and/or crypto-related.}
\label{tab:topliterature}
{\resizebox{\textwidth}{!}{%
\begin{tabular}{ R{3.5cm} | p{0.45\linewidth} c c c c}
\toprule
\rotatebox[origin=l]{0}{Reference} & \rotatebox[origin=l]{0}{ Title}& \rotatebox[origin=l]{0}{ Year} & \rotatebox[origin=l]{0}{ Citations} & \rotatebox[origin=l]{0}{ Review} & \rotatebox[origin=l]{0}{ Crypto}\\
\midrule
\citet{NGAI2011559} & The application of data mining techniques in financial fraud detection: A classification framework and an academic review of literature&2011& 1307& \ding{51} & \\
\rowcolor{gray!10}\citet{FRONZETTICOLLADON201749} & Using social network analysis to prevent money laundering&2017& 200&   & \\
\citet{chang2007wirevis} & WireVis: Visualization of Categorical, Time-Varying Data From Financial Transactions&2007& 171&   & \\
\rowcolor{gray!10}\citet{senator1995financial} & Financial Crimes Enforcement Network AI System (FAIS) Identifying Potential Money Laundering from Reports of Large Cash Transactions&1995& 152&   & \\
\citet{DREZEWSKI201518} & The application of social network analysis algorithms in a system supporting money laundering detection&2015& 141&   & \\
\rowcolor{gray!10}\citet{mcginn2016visualizing} & Visualizing Dynamic Bitcoin Transaction Patterns&2016& 122&   &\ding{51}\\
\citet{gao2007framework} & A framework for data mining-based anti-money laundering research&2007& 106& \ding{51} & \\
\rowcolor{gray!10}\citet{wu2021detecting} & Detecting Mixing Services via Mining Bitcoin Transaction Network With Hybrid Motifs&2021& 94&   &\ding{51}\\
\citet{malm2013using} & Using friends for money: the positional importance of money-launderers in organized crime&2013& 66&   & \\
\rowcolor{gray!10}\citet{10.1145/3097983.3098015} & A Local Algorithm for Structure-Preserving Graph Cut&2017& 63&   & \\
\citet{li2020flowscope} & FlowScope: Spotting Money Laundering Based on Graphs&2020& 54&   & \\
\rowcolor{gray!10}\citet{irwin2018illicit} & Illicit Bitcoin transactions: challenges in getting to the who, what, when and where&2018& 47& \ding{51} &\ding{51}\\
\citet{phetsouvanh2018egret} & EGRET: Extortion Graph Exploration Techniques in the Bitcoin Network&2018& 38&   &\ding{51}\\
\rowcolor{gray!10}\citet{kute2021deep} & Deep Learning and Explainable Artificial Intelligence Techniques Applied for Detecting Money Laundering–A Critical Review&2021& 38& \ding{51} & \\
\citet{zhdanova2014no} & No Smurfs: Revealing Fraud Chains in Mobile Money Transfers&2014& 32&   & \\
\rowcolor{gray!10}\citet{bhalerao2019mapping} & Mapping the underground: Supervised discovery of cybercrime supply chains&2019& 24&   & \\
\citet{ovelgonne2012covertness} & Covertness Centrality in Networks&2012& 20&   & \\
\rowcolor{gray!10}\citet{cheong2010event} & Event-based approach to money laundering data analysis and visualization&2010& 12&   & \\
\citet{gerbrands2022effect} & The effect of anti-money laundering policies: an empirical network analysis&2022& 11& \ding{51} & \\
\rowcolor{gray!10}\citet{zhou2022behavior} & Behavior-Aware Account De-Anonymization on Ethereum Interaction Graph&2022& 7&   &\ding{51}\\
\bottomrule
\end{tabular}%
}}
{}
\end{table}

\subsubsection{General Classification.}
Five out of 11 review papers are selected. 
This is to be expected, since review papers have a higher likelihood of being cited~\citep{dong2005impact}. 
Plots summarising the classification according to Table~\ref{tab:classification} are given in Appendix~B. 
Similar results are observed for the top-cited papers as for the literature as a whole; unsupervised learning is by far the most applied method, and research relies heavily on manual and rule-based features. 
Only time and threshold-dependent metrics were used for evaluation. 
This indicates that threshold-independent metrics are not mainstream when comparing AML models using networks analytics. 
We observe that, compared to the wider literature, flow/chain detection methods are better cited as well as studies using open-source data. 
As mentioned before, flow detection better captures the intricate patterns used in money laundering. 
The use of open-source data makes comparison with and replication of the studies possible, fostering research adoption.

Additionally, five papers deal with crypto-related money laundering, indicating a strong research interest. 
On the one hand, crypto currencies are widely used by criminals because of the absence of regulation~\citep{fletcher2021countering} and the anonymity it provides~\citep{irwin2018illicit, weber2019anti}. 
On the other hand, a full history of ownership is available on the blockchain for each \textit{coin}, facilitating the discovery of transaction patterns in payment networks~\citep{irwin2018illicit, weber2019anti}. 

\subsubsection{Task-Level Objectives.}
Table~\ref{tab:summary objectives} summarises the data set used with information therein, the model proposed by the authors to capture a certain money laundering phenomenon, and the baselines to which the proposed model is compared. 
Additionally, Table~\ref{tab:network} summarises the network properties of these papers. 

\begin{table}[t]
	\centering
	\caption{Summary of the objective of the paper, where the paper uses [Data set] with [information] to introduce a [model] capturing [phenomenon] and is compared to [baselines].}
    \label{tab:summary objectives}
	{\resizebox{\textwidth}{!}{%
			\begin{tabular}{L{4cm}|L{4cm}L{4cm}L{4cm}L{4cm}L{4cm}}
				\toprule
				Paper & Data set & Information & Model & Phenomenon & Baselines \\ \midrule
				\citet{FRONZETTICOLLADON201749} & Italian factoring data & Transaction~invoices, economic~sector, geographical~area & Correlation of centrality metrics with label & Smurfing & - \\
				 \rowcolor{gray!10}\citet{chang2007wirevis} & Proprietary bank data set & Transactions, accounts, keywords, personal~information & Visualisation of keywords-accounts, account using transactions, and evolution transactions of (groups of) accounts & Suspicious behaviour & - \\
				\citet{senator1995financial} & Reports of large cash transactions according to Bank Secrecy Act & People, accounts, businesses, locations and transactions & Graphical user interface & identifying high-value leads & - \\
				 \rowcolor{gray!10}\citet{DREZEWSKI201518} & Proprietary bank statements, national court register & Connections & Rule-based role definition via centrality metrics & Roles in criminal organisations & - \\
				\citet{mcginn2016visualizing} & Bitcoin & Transactions, wallets & Visualisation & Unexpected high-frequency transaction patterns & - \\
				 \rowcolor{gray!10}\citet{wu2021detecting} & Bitcoin & Transactions and wallets & Account, transaction and network feature extraction + z-score based on hybrid motives + logistic regression & Mixing & - \\
				\citet{malm2013using} & Police intelligence reports & Person under investigation in the report and relations, demographic characteristics of each person, nature of illicit drug trade involvement, and the types of relationships that existed among individuals & Correlation of betweenness and eigen-vector centrality with known roles & Role of money-launderers in illicit markets & - \\
				 \rowcolor{gray!10}\citet{10.1145/3097983.3098015} & Proprietary bank data set & Bank accounts, names, emails, addresses, and phone number & HOSPLOC (local graph clustering) & Synthetic identities and money laundering & Local clustering (Nibble, NPR, LS-OSC), global (NMF, TSC) \\
				\citet{li2020flowscope} & Proprietary and Czech Financial Data set & Bank accounts, transactions  & FlowScope (objective maximisation to find fast money flow through bank in multi-partite network) & Anomalous money flow indicative for money laundering & SpokEn, D-Cube, Fraudar, HoloScope, RRCF \\
				 \rowcolor{gray!10}\citet{phetsouvanh2018egret} & Bitcoin blockchain and Ashley Madison blackmail & Transactions and wallets & Average length and confluence analysis & Suspicious bitcoin flow and discover other wallets owned by suspected perpetrators  & - \\
				\citet{zhdanova2014no} & Synthetic MMT data & Transaction between phones & PSA@R (event-driven process analysis) & Smurfing & PART, C4.5, Random Forest \\
				 \rowcolor{gray!10}\citet{bhalerao2019mapping} & Forum threads (English and Russian) & Users, Posts and replies & Automatic classification using XGBoost, LogReg, SVM ans FastText. Construction supply chain based on interaction graph. Visualisation via alluvial plots & Money laundering and other fraudulent behaviour & - \\
				\citet{ovelgonne2012covertness} & Youtube and Universitat Rovira I Virgili data sets & Network & Covertness centrality & Hiding behaviour & - \\
				 \rowcolor{gray!10}\citet{cheong2010event} & Criminal relations database & Transactions and relationships & Rule-based degree of suspicion and visualisation of relations & Find additional suspects in money laundering cases & - \\
				\citet{gerbrands2022effect} & iCOV & Transactions, family and professional relations of people investigated & Clustering (Louvain and temporal) + node level centralities & Interactions over time & - \\
				 \rowcolor{gray!10}\citet{zhou2022behavior} & Ethereum (Eth-ICO, Eth-Mining, Eth-Exchange, Eth-Pish\&Hack) &  Transactions and wallets & Pipeline based on sampling, Hierarchical Graph ATtention Encoder and sub-graph contrastive learning & Account identification (phishing, exchange etc.) & Manual, Deepwalk, node2vec, trans2vec, graph2vec + LR, RF, LGBM, and GCN, GAT, GIN, I2BGNN-A, I2BGNN-T \\
				\bottomrule
			\end{tabular}%
	}}
	{}
\end{table}

We again see a strong reliance on \textit{basic} centrality measures for anti-money laundering. 
A couple of papers apply very basic analysis. 
\citet{FRONZETTICOLLADON201749} and \citet{malm2013using} purely consider correlations to extract conclusions for their AML features. 

Next to the typical objectives of money laundering detection, some research illustrates how network analytics can be used to support investigators to find novel leads and other people involved in previously-reported cases.

Table~\ref{tab:summary objectives} points to limitations in the literature. 
Many of the proposed methods are only tested on proprietary data, limiting the reproducibility of their results. 
The performance is often illustrated using case studies, without the inclusion of other baselines. 
This shows that there is a lack of a consensus on baselines and state-of-the-art methods in the field. 

The above observations motivate the construction of the experimental set-up of Section~\ref{sec:benchmark}. 
Our aim is to come to a common framework to test methods for network analytics in a uniform way.

In addition to the objectives of the studies, Table~\ref{tab:network} presents the network construction in these papers. 
It indicates that most of the papers use homogeneous networks with transactions forming the edges between accounts. 
Only a few use heterogeneous networks, either by having both accounts and transactions as nodes, or having nodes that represent different characteristics or feature values.

\begin{table}[t]
    \centering
    \caption{Summary of the construction of the network.}
    \label{tab:network}
    {\begin{tabular}{L{4cm}|L{6cm}L{5.5cm}}
    \toprule
    \textbf{Paper} & \textbf{Nodes} & \textbf{Edges} \\ \midrule
    \citet{FRONZETTICOLLADON201749} & Factoring companies &\begin{tabular}{@{}l} Economic sector \\ geographical data \\ transaction amount \\ tacit link (same owners/resources) \end{tabular} \\
    \rowcolor{gray!10}\citet{chang2007wirevis} & People & Transactions \\
    \citet{senator1995financial} & \begin{tabular}{@{}l} Accounts \\ People, businesses, accounts and location
    \end{tabular} & Connections \\
    \rowcolor{gray!10}\citet{DREZEWSKI201518} & People (via court register) & Transactions (via bank statements) \\
    \citet{mcginn2016visualizing} & Transactions, inputs and outputs & Input/output part of same transaction, inputs that belong to the same address \\
    \rowcolor{gray!10}\citet{wu2021detecting} & \begin{tabular}{@{}l} Address \\ Address and transactions
    \end{tabular} & \begin{tabular}{@{}l} Transactions \\ If they are related
    \end{tabular} \\
    \citet{malm2013using} & People (from police reports) & Co-appearance in the report (family, client-lawyer etc.)\\ 
    \rowcolor{gray!10}\citet{10.1145/3097983.3098015} & All items of interest (bank account, name, email, address, phone number) & Link bank account to other items \\
    \citet{li2020flowscope} & Bank accounts & Transfers \\
    \rowcolor{gray!10}\citet{phetsouvanh2018egret} & Bitcoin wallet address & Transactions \\
    \citet{ovelgonne2012covertness} & - & - \\
    \rowcolor{gray!10}\citet{cheong2010event} & People & Relations \\
    \citet{zhou2022behavior} & Accounts & Transactions \\
    \rowcolor{gray!10}\citet{zhdanova2014no} & mWallets & Transactions \\ 
    \bottomrule
    \end{tabular}}
     {}
\end{table}

\subsubsection{Processing steps.}
Next, the papers are classified according to the framework presented by~\citet{bockel2023fraud}.
Table~\ref{tab:processing} summarises the papers over the processing steps. 
The most prevalent pre-processing steps are feature engineering and exploration. 
Feature engineering is done to define different risk profiles or to enhance GNN models. 
This is either based on neighbourhood and centrality metrics~\citep{FRONZETTICOLLADON201749, DREZEWSKI201518, ovelgonne2012covertness} or on the aggregation of local (node-specific) information~\citep{zhou2022behavior, zhdanova2014no}. 
For exploration, some research is done on how to construct meaningful visualisations~\citep{chang2007wirevis, mcginn2016visualizing}, while other papers rely on clustering the nodes as additional support for their findings~\citep{malm2013using, FRONZETTICOLLADON201749}.

Most papers apply unsupervised learning in their processing step. 
This involves detection of suspicious transaction flows~\citep{10.1145/3097983.3098015, li2020flowscope, phetsouvanh2018egret}. 
Another important part is the use of centrality metrics to define specific roles for entities in the network~\citep{DREZEWSKI201518, malm2013using, ovelgonne2012covertness}. 
Finally, some methods create heuristics to define outlying behaviour~\citep{senator1995financial, chang2007wirevis, cheong2010event, zhdanova2014no}.

For the post-processing step, most research covers statistical evaluation, implementation and interpretation. 
Statistical evaluation is used to obtain a quantitative evaluation of the performance. 
Implementation consists of analysing the efficiency of the methods. 
It deals solely with computation time and scalability. 
For interpretation, researchers use specific interpretable features. 

\begin{table}[t]
    \centering
    \caption{Papers per method cluster, based on~\citet{bockel2023fraud}.}
    \label{tab:processing}
    {\begin{tabular}{l|lC{6cm}}
    \toprule
         \textbf{Process Step} & \textbf{Sub-Steps} & \textbf{References} \\ \midrule
         \multirow{7}{*}{Pre-processing}& Sampling &\citep{FRONZETTICOLLADON201749}.\\
         & Exploration & \citep{FRONZETTICOLLADON201749},\citep{chang2007wirevis},\citep{mcginn2016visualizing}, \citep{malm2013using}.\\
         & Missing Value Treatment & \citep{FRONZETTICOLLADON201749}. \\
         & Outlier Detection and Treatment & - \\
         & Categorization, standardization \& segmentation & \citep{DREZEWSKI201518}, \citep{bhalerao2019mapping}. \\
         & Feature Engineering  & \citep{FRONZETTICOLLADON201749}, \citep{DREZEWSKI201518}, \citep{ovelgonne2012covertness},\citep{zhou2022behavior}, \citep{zhdanova2014no}. \\
         & Variable Selection  & - \\
         \midrule
         \multirow{4}{*}{Processing} & Unsupervised Learning & \citep{chang2007wirevis}, \citep{senator1995financial}, \citep{DREZEWSKI201518}, \citep{malm2013using}, \citep{10.1145/3097983.3098015}, \citep{li2020flowscope}, \citep{phetsouvanh2018egret}, \citep{ovelgonne2012covertness}, \citep{cheong2010event},\citep{zhdanova2014no}. \\
         & Supervised Learning & \citep{FRONZETTICOLLADON201749}, \citep{bhalerao2019mapping}. \\
         & Semi-Supervised Learning & \citep{wu2021detecting}. \\
         & Hybrid Learning & \citep{zhou2022behavior}. \\
         \midrule
         \multirow{4}{*}{Post-processing} & Statistical Evaluation & \citep{FRONZETTICOLLADON201749}, \citep{wu2021detecting}, \citep{10.1145/3097983.3098015}, \citep{li2020flowscope}, \citep{bhalerao2019mapping}, \citep{zhou2022behavior}, \citep{zhdanova2014no}. \\ 
         & Interpretation & \citep{DREZEWSKI201518}, \citep{wu2021detecting}, \citep{bhalerao2019mapping}, \citep{zhou2022behavior}. \\ 
         & Economical Evaluation & \citep{zhdanova2014no}. \\ 
         & Implementation & \citep{DREZEWSKI201518}, \citep{mcginn2016visualizing}, \citep{10.1145/3097983.3098015}, \citep{li2020flowscope}, \citep{ovelgonne2012covertness},\citep{zhou2022behavior}, \citep{zhdanova2014no}. \\ \bottomrule
    \end{tabular}}
     {}
\end{table}

More detailed trends in the literature are discussed below according to three main categories, that mainly follow the processing steps; unsupervised learning, (semi-)supervised learning and visualisation methods. 

\textbf{Unsupervised learning} is divided into two main streams. 
The first assigns and identifies specific roles of actors in criminal networks, based on police reports of criminal cases~\citep{ malm2013using, DREZEWSKI201518}. 
The second covers methods to track money flows through the network~\citep{10.1145/3097983.3098015, li2020flowscope, phetsouvanh2018egret}. 
It is important to integrate these transaction chains, since money laundering is a process involving multiple steps and actors~\citep{levi2006money}. 
Next to these two main streams, the remaining paper~\citep{ovelgonne2012covertness} introduces a new covertness measure. 

Although detecting money laundering is in principle trying to find anomalous behaviour, very few studies deal with (unsupervised) anomaly detection~\citep{magomedov2018anomaly,prado2017community,dumitrescu2022anomaly}, leaving possibilities for future work. 

\textbf{(Semi-)Supervised learning} methods deal with different aspects of money laundering. 
Their strength lies in a combination of (1) the methods used and (2) tailoring the network to the problem at hand. 
The obtained network features are put into downstream classifier. 
To maintain interpretability, network metrics are recalculated for multiple networks, each representing a specific aspect of the data, to extract predictive features~\citep{FRONZETTICOLLADON201749, wu2021detecting}. 
Promising methods are based on deep learning~\citep{zhou2022behavior}, but these are mostly black boxes. 
Research on the interpretability of deep network representation methods is still scarce~\citep{kute2021deep}.

\textbf{Visualisation} of transaction networks is developed to support the experts in during investigations. 
This is either done by having a overview of the flow of money over different transactions~\citep{mcginn2016visualizing, chang2007wirevis} or by finding specific relations among the persons involved~\citep{senator1995financial, didimo2011advanced, cheong2010event}. 
Hence, these methods are mostly intended to be used after suspicion has been raised. 
None of them include an evaluation of the added value in money laundering detection. 

\subsubsection{Challenges.}
Table~\ref{tab:challenges} summarises the papers according to the challenges identified by~\citet{bockel2023fraud}. 
It only contains those challenges mentioned in the literature. 
The main challenge is data availability, followed by unlabelled data. 
Due to privacy issues, only a limited number of anonymised data is available. 
Additionally, since transaction data sets are often huge and resources are limited, it is hard to provide all instances with a label. 

Few papers mention feature construction, real-time execution, or verification latency as a challenge. 
Anti-money laundering is less reliant on execution time, since the investigation is done after the fact, based on a longer history of transactions~\citep{levi2006money}. 

We introduce four additional challenges observed in the anti-money laundering literature, supplementing~\citet{bockel2023fraud}. 
We view this as an extension of the framework since these challenges can be important to fraud research in general.
\begin{itemize}
    \item \textbf{Bias in data} affects the performance of the model and manifests itself in different ways. First, there is \textit{bias in the missing values}~\citep{malm2013using}. Second, \textit{detection bias}~\citep{zhdanova2014no} can be present in synthetic data when specific patterns are present that the model is trained to look for.
    \item \textbf{Generalisation} is mainly used in two ways. One way refers to the ability for methods developed for anti-money laundering to be used for other fraud domains as well~\citep{zhou2022behavior}. Another and important use of the term refers to the need for methods to generalise to unseen data~\citep{goodfellow2016deep}, often referred to as \textit{inductive} methods (compared to \textit{transductive} ones). This is an important consideration when selecting network-based methods~\citep{hamilton2017inductive, van2020representation, van2022inductive}.
    \item \textbf{Robustness} of a method is important since money laundering can be seen as an adversarial attack~\citep{li2020flowscope}, as perpetrators try to avoid detection by the methods. 
    \item \textbf{Anonymity} of the account data gives rise to two main challenges. First, the identity of the account holder is unknown in crypto data sets, which makes it almost impossible to know if different accounts - also called wallets - belong to the same person~\citep{wu2021detecting}. Second, financial institution possess customer data, but due to privacy reasons, researchers cannot publish these. Hence, published articles avoid using Personally Identifiable Information all together~\citep{bhalerao2019mapping}.
\end{itemize}

\begin{table}[t]
    \centering
    \caption{Papers per fraud detection challenge, based on~\citet{bockel2023fraud}.}
    \label{tab:challenges}
    {\begin{tabular}{l|lC{10cm}}
    \toprule
          & \textbf{Challenge} & \textbf{Papers} \\ \midrule
         \multirow{11}{*}{Original}& Automation & \citep{senator1995financial}, \citep{cheong2010event}, \citep{zhou2022behavior}. \\
         & Class Imbalance & \citep{wu2021detecting}, \citep{bhalerao2019mapping}. \\
         & Concept Drift & \citep{FRONZETTICOLLADON201749}, \citep{bhalerao2019mapping}. \\
         & Data Availability & \citep{FRONZETTICOLLADON201749}, \citep{chang2007wirevis}, \citep{wu2021detecting}, \citep{malm2013using}, \citep{ovelgonne2012covertness}, \citep{zhdanova2014no}. \\
         & Feature Construction & \citep{ovelgonne2012covertness}. \\
         & Noisy Data & \citep{senator1995financial}, \citep{wu2021detecting}, \citep{malm2013using}. \\
         & Stream Data & \citep{senator1995financial}, \citep{mcginn2016visualizing}, \citep{zhou2022behavior}. \\
         & Real-Time Execution & \citep{FRONZETTICOLLADON201749}. \\
         & Scalability & \citep{10.1145/3097983.3098015}, \citep{ovelgonne2012covertness}, \citep{zhou2022behavior}. \\
         & Unlabelled Data & \citep{senator1995financial}, \citep{wu2021detecting}, \citep{bhalerao2019mapping}, \citep{zhou2022behavior}. \\
         & Verification Latency & \citep{zhdanova2014no}. \\
         \midrule
         \multirow{4}{*}{New} & Bias in data & \citep{malm2013using}, \citep{zhdanova2014no}. \\
         & Generalisation & \citep{zhou2022behavior}. \\
         & Robustness & \citep{li2020flowscope}. \\
         & Anonymity & \citep{mcginn2016visualizing}, \citep{wu2021detecting}, \citep{bhalerao2019mapping}. \\  \bottomrule
    \end{tabular}}
    {}
\end{table}

\subsection{Further Considerations}
\label{subsec:furth considerations}
Based on the observations made above, we give an intermediate conclusion and perspective on future work. 
The methods in the literature still rely heavily on expert-based methods, i.e., manual feature engineering and rule-based systems, but deep learning methods are gaining traction. 
A more detailed analysis of these methods is needed to evaluate their performance properly, especially given that the current literature barely applies threshold-independent metrics meant for highly skewed data, like the AUC-PR. Moreover, the top-cited papers barely compare their methods to other baselines. 
This detailed analysis is the topic for the second part of this paper.

Different gaps in the literature exist. 
A first gap was also identified by~\citet{kute2021deep}, namely the lack of interpretation tools. 
Future work should interpretation methods to try and explain the output of graph neural networks.

A second gap concerns the under-representation of classic unsupervised methods~\citep{chandola2009anomaly, akoglu2015graph}. 
Future work should analyse the effectiveness of unsupervised learning in networks to find anomalous behaviour. 
This can be beneficial since many clients and transactions are effectively unlabelled, and the modi operandi for money laundering keep evolving. 
The second part of this paper will introduce a first way of applying such unsupervised learning methods, but a more extensive study is needed.

A third is that most research studies static networks. 
Some research is being done to incorporate the time-dynamic nature of transaction networks using recurrent neural networks, but we believe that much more can still be done.

A fourth gap is related to the inherent incompleteness of labels in AML. 
As criminals try to cover their tracks and financial institutions have only limited resources~\citep{van2023catchm,DBLP:journals/corr/abs-1009-6119}, it is a given that there are still many money laundering cases that stay unnoticed. 
These transactions are seen as having label 0 when training the models, resulting in a positive and unlabelled (PU) problem. 
Specific PU-learning methods for networks should be applied when training a model. 

A fifth gap is related to the limited resources as well. 
Since not all clients and transactions can be investigated, the predictions should take this into account. 
One way of dealing with this is learning-to-rank~\citep{devriendt2022LTR,VANDERSCHUEREN2024114151}, which has already been shown to be effective in fraud detection. 
The main benefit is that the goals is to correctly rank the most suspicious clients, while less importance is given to have a correct ranking for less suspicious clients. Although this is highly relevant, it has not yet been applied to the AML literature using network analytics.

\section{Experimental Set-Up}
\label{sec:benchmark}
To the best of our knowledge, a benchmark study comparing the state-of-the-art network learning methods for anti-money laundering is still missing. 
This section describes the experimental framework to fill this gap, including model specification (Section~\ref{subsec:model}), data (Section~\ref{subsec:data}), hyperparameter tuning (Section~\ref{subsec:hyperparam}) and performance metrics (Section~\ref{subsec:metrics}). 
The focus is on supervised learning, but unsupervised learning is also explored. 
Given challenges in open-source data availability, we present results on two data sets and provide code to facilitate reproducible experiments.
We hope this framework will support and expand future research, and allows uniform publishing and comparing of results on proprietary data without the need to disclose sensitive information.

\subsection{Model Specification}
\label{subsec:model}
To set up the experimental evaluation, we start from the wider literature on network representation. 
Extensive overviews are given by~\citet{cai2018comprehensive, hamilton2017inductive} and \citet{goyal2018graph}. 
A sub-branch of representation learning applies deep learning methods~\citep{zhou2020graph,wu2020comprehensive}. 
Based on these papers, we select the methods that are most prominently used. 

We refined the selection by considering that, for AML, transaction data contains networks with millions of nodes and edges, making scalability an important selection criterion. 
Table~\ref{tab:features} provides an overview of the methods selected. 
They are grouped into three categories~\citep{cai2018comprehensive, hamilton2017inductive,goyal2018graph}:
manual feature engineering, shallow representation learning and deep representation learning. 
These methods will be tested against a baseline model that only includes the intrinsic features, denoted by IF.

\begin{table}[t]
    \centering
    \caption{The methods used to generate network features for the experimental evaluation.} 
    \label{tab:features}
    {\begin{tabular}{l|lL{8cm}}
    \toprule
         \textbf{Category} & \textbf{Features} & \textbf{Definition}  \\ \midrule
        \multirow{3}{*}{Manual feature engineering}         & Density & \citet{baesens2015fraud}: Number of edges in the egonet relative to the maximal number possible.   \\ 
                                                            & Centrality measures & Closeness, betweenness and eigenvector centrality. \\
                                                            & PageRank & \citet{page1999pagerank}: Score indicating the importance of a node. \\ \midrule
        \multirow{2}{*}{Shallow representation learning}    & DeepWalk & \citet{perozzi2014deepwalk}: Random walk through the network. Embedding via word2vec. \\
                                                            & Node2vec & \citet{grover2016node2vec}: Truncated random walks (breadth-first vs. depth-first). Embedding via word2vec. \\
                                                            \midrule
        \multirow{4}{*}{Deep representation learning}       & GCN & \citet{kipf2017semisupervised}: Graph convolutional network \\
                                                            & GraphSAGE & \citet{hamilton2017inductive}: Aggregation based on fixed-size sample of neighbours \\
                                                            & GAT & \citet{veličković2018graph}: Graph attention network \\
                                                            & GIN & \citet{xu2019powerful}: Graph isomorphism network \\ \bottomrule
    \end{tabular}}
    {}
\end{table}

\textbf{Manual feature engineering} includes both local and global metrics. 
The first is the density of the ego-network. 
A node's ego-network consists of the node, its direct neighbours and all connections between the neighbours. 
The density is the relative number of connections in this ego-network compared to the theoretical maximum. 
Summary statistics, i.e., minimum, mean and maximum values, of the density of the node's neighbours are also included. 

A second type consists of centrality metrics, quantifying the global position of the node in the network~\citep{10.1093/acprof:oso/9780199206650.001.0001}. 
The most popular centrality metrics are the betweenness, closeness and eigenvector centrality. 
The betweenness quantifies the number of shortest paths between any two nodes in the network passing through that node. 
The closeness, being the inverse of the average distance, measures how close a node is to all other nodes. 
The eigenvector centrality is high for nodes that are connected to important nodes in the network. 
The importance of the node itself is quantified using the PageRank~\citep{page1999pagerank}. 

Next to the manual feature engineering, automatic network features are constructed using network embedding methods~\citep{hamilton2017inductive, goyal2018graph, cai2018comprehensive}. 
Although there are many different methods to construct a network embedding~\citep{cai2018comprehensive}, we use those based on deep learning, called network representation learning~\citep{van2020representation,van2023catchm}. 

\textbf{Shallow network representation learning} can be viewed as an ``embedding lookup''~\citep{hamilton2017inductive, van2023catchm}. 
We apply methods based on random walks. 
DeepWalk~\citep{perozzi2014deepwalk} was the first method to try to give nodes that are close in the network, i.e., that co-occur on short random walks, a similar embedding. 
Each random walk is seen as a sentence of words. 
The nodes are embedded into a Euclidean latent space using NLP methods, often skip-gram~\citep{mikolov2013efficient}. 

Node2vec~\citep{grover2016node2vec} extends DeepWalk by introducing two hyperparameters, $p$ and $q$.
The method samples each node according to the following unnormalized transition probability, $\alpha$, at node $v$, when coming from node $t$:
\[ \alpha(c_{i+1} = x\mid c_i = v, c_{i-1} = t) = \begin{cases}
    \frac{1}{p} & \text{if } d_{tx} = 0 \\
    1 & \text{if } d_{tx} = 1 \\
    \frac{1}{q} & \text{if } d_{tx} = 2\\
\end{cases}, \]
where $d_{tx}$ denotes the distance between node $t$ and $x$. 
These probabilities are defined a link exists between $v$ and $x$. 
Otherwise, it is set to 0.

The main drawbacks of shallow representation learning methods are that (1) they are transductive, so need full retraining when a new node is added to the network, and (2) retraining on the same network results in a different embedding. 
This second point can be understood intuitively by viewing two embeddings of the same network, with one being a rotation of the other. 
Both are equally good, but a downstream classifier trained on one cannot be used to make predictions on the other, as the coordinates of the nodes are different.

In the experiments below, we construct the embedding on all nodes in the training and validation set, while only training on the labels from the training set. 
Afterwards, the embedding is constructed for the full network, and we train a new classifier using the labels of the train and validation set.

This mimics the applications of these methods in reality, since a bank will have the full transaction network up to the current time, with some historical labels. 
Based on this, they need to determine which accounts to investigate next for money laundering. 

\textbf{Deep network representation learning} adapts deep learning methods to be directly applicable on networks. 
We apply different graph neural network~(GNN) architectures. 

The most popular methods for node embedding calculation are graph convolutional networks (GCNs), based on convolutional neural networks~\citep{kipf2017semisupervised}. 
It relies on neighbourhood aggregation to update the embedding as follows:
\[ H^{(l+1)} = \sigma\left( \tilde{D}^{-1/2} \tilde{A} \tilde{D}^{-1/2} H^{(l)} W^{(l)} \right), \]
with $H^{(l)}$ the embedding at step $l$, $\sigma$ a non-linear (activation) function, $\tilde{A} = A+I$ the adjacency matrix with self-loops added, the diagonal matrix $\tilde{D}_{ii} = \sum_j \tilde{A}_{ij}$, and $W^{(l)}$ the learnable weights in layer $l$. 
The node embedding for the individual node is updated via \[ h_i^{(l+1)} = \sigma\left(\sum_{j\in\tilde{\mathcal{N}}_i} \frac{1}{\sqrt{\text{deg}(i)\text{deg}(j)}} W^{(l)} h_j^{(l)} \right). \]

Neighbourhood aggregation has seen multiple extensions, two of which this paper will evaluate. 
The first is GraphSAGE~\citep{hamilton2017inductive}, which samples a fixed number of neighbours for each node, to make the calculations scalable to large graphs. 
The representations of the sampled neighbours are aggregated, with the authors proposing the mean, LSTM and pooling aggregators~\citep{hamilton2017inductive}.
Then, the aggregated neighbourhood representation is concatenated with the node's representation. 
Finally, the representation is updated using the weights and activation function. 
For a given node $v$, this is calculated as: 
\begin{eqnarray*}
    h_{\mathcal{N}_v}^{(l)} &=& \texttt{AGGREGATE}_l\left(\{h_u^{(l-1)}, \forall u \in \mathcal{N}_v\}\right)\\
    h_v^{(l)} & = & \sigma\left(W^{(l)}\cdot \left[h_v^{(l-1)} \mathbin\Vert h_{\mathcal{N}_v}^{(l)}\right] \right),
\end{eqnarray*} where $\mathbin\Vert$ represents the concatenation operator. 

The second extension is graph attention network (GAT)~\citep{veličković2018graph}, which adds an attention mechanism. 
Our work uses the modified version of the attention mechanism, introduced to fix the static attention problem~\citep{DBLP:journals/corr/abs-2105-14491}:
\begin{eqnarray*}
    e_{ij} & = &a^T \texttt{LeakyReLU}\left(W^{(l)} \left[h_v^{(l-1)} \mathbin\Vert h_{\mathcal{N}_v}^{(l)}\right] \right) \\
    \alpha_{ij} &=& \texttt{softmax}_j(e_{ij}) \\
    h_i^{(l)} &=& \sigma\left( \sum_{j\in\tilde{\mathcal{N}}_i} \alpha_{ij} W^{(l)} h_j^{(l-1)}. \right)
\end{eqnarray*}
As mentioned by~\citet{veličković2018graph}, having multiple mechanisms---also called heads---can improve stability. 
We follow \citet{veličković2018graph}, where the intermediate steps concatenate the results over the heads, while the final layer averages the results. 

The final method is the Graph Isomorphism Network (GIN)~\citep{xu2019powerful}. 
The authors used the Weisfeiler-Lehman graph isomorphism test to define the necessary conditions a GNN must satisfy to achieve maximal discriminative power. 
This discriminative power is denoted by how well the training data is fitted. 

Using the universal approximation theorem~\citep{hornik1989multilayer, hornik1991approximation}, GIN updates the representation using a multilayer perceptron~(MLP). 
\[ h_v^{(l)} = \texttt{MLP}^{(l)} \left( \left(1+\epsilon^{(l)}\right) h_v^{(l-1)} + \sum_{j\in\mathcal{N}_i}h_j^{(l-1)} \right) \]
Although this increases the risk of overfitting, the authors show that GIN has satisfactory generalisation capabilities. 
The main problem with other GNNs is that they tend to underfit the data~\citep{xu2019powerful}. 
Recent research showed GIN's potential to outperform other GNN architectures~\citep{egressy2024provably,tiukhova2023inflectdgnn}.

There are some key differences between the three categories presented in Table~\ref{tab:features}. 
The manual features need to be specifically defined, while representation learning tries to find meaningful embeddings automatically. 
Additionally, the manual features and the shallow representations are added to the intrinsic features, which are used in a down-stream classifier (as illustrated in Figure~\ref{fig:embedding illustration}). 
The GNNs incorporate the intrinsic features and directly learn the embeddings on the classification task. 

\begin{figure*}
    \centering
    {\begin{tikzpicture}[node distance=1cm, >=Stealth, every node/.style={font=\footnotesize}]
			\node[minimum width=4cm, minimum height=3cm] (black_table) {
				\begin{tabular}{|c|cccc|}
					\hline
					\textbf{ID} & \textbf{F1} & \textbf{F2} & $\hdots$ & \textbf{Fn} \\
					\hline
					A & f11 & f21 & $\hdots$ & fn1 \\
					B & f12 & f22 & $\hdots$ & fn2 \\
                    $\vdots$ & $\vdots$ & $\vdots$ & $\ddots$ & $\vdots$ \\
					E & f1k & f2 & $\hdots$ & fnm \\
					\hline
				\end{tabular}
			};
			\node[minimum width=4cm, minimum height=3cm, right=of black_table] (blue_table) {
				\begin{tabular}{|c|cccc| >{\columncolor{blue!20}}c >{\columncolor{blue!20}}c>{\columncolor{blue!20}}c|}
					\hline
					\textbf{ID}& \textbf{F1} & \textbf{F2} & $\hdots$ & \textbf{Fn} & \textbf{NF1} & $\hdots$ &\textbf{NFm} \\
					\hline
					A & f11 & f21 & $\hdots$ & fn1 & nf11 & $\hdots$ & nfm1 \\
					B & f12 & f22 & $\hdots$ & fn2 & nf12 & $\hdots$ & nfm2 \\
                    $\vdots$ & $\vdots$ & $\vdots$ & $\ddots$ & $\vdots$ & $\vdots$ & $\ddots$ & $\vdots$ \\
					E & f1k & f2 & $\hdots$ & fnm & nf1k & $\hdots$ & nfmk\\
					\hline
				\end{tabular}
			};
			\draw[->, thick] (black_table) -- (blue_table) node[midway, above, font=\small]{};
			\coordinate[above=of black_table] (network_pos);
			\begin{scope}[shift={(network_pos)}]
				\foreach \x/\y/\label in {0/0/A, 1.5/1/B, 3/0/C, 1.5/-1/D, 1.5/0/E}
				\node[circle, draw, thick, fill=blue!40, inner sep=1.5pt] (\label) at (\x,\y) {\label};
				\foreach \source/\dest in {A/B, B/C, C/D, D/E, E/A, E/B}
				\draw[thick] (\source) -- (\dest);
			\end{scope}
			\draw[->, thick] (C) to[bend left=45] node[midway, above=1em, font=\small]{Network Embedding} (blue_table);
		\end{tikzpicture}}
    \caption{For the feature engineering and shallow representation methods, the obtained network features/embedding is added to the intrinsic ones.}
    \label{fig:embedding illustration}
    {}
\end{figure*}

Different decoders are used for the final classification in the supervised setting. 
A neural network with two hidden layers, each of dimension ten is used for the manual features and shallow representation methods. 
For the graph neural networks, the GNN layers are followed by a single linear layer for making predictions. 

For the unsupervised model, we apply isolation forest~\citep{lui2012isolation}, as it achieves state-of-the-art performance for outlier/anomaly detection~\citep{tiukhova2022Ooutlierdetection, jia2020anomaly}. 
This will be applied to the intrinsic and manual features, and the shallow representations, since it requires a tabular input. 
A larger experiment comparing unsupervised methods for network analytics is left for future work.

One of the objectives of this research is to provide the code for all methods in a clean and simple way and make these methods easily accessible to researchers and practitioners. 
The methods are implemented in Python. 
The manual features are constructed using NetworkX~\citep{SciPyProceedings_11} and NetworKit~\citep{Angriman2022}, and the representation learning is implemented in PyTorch Geometric~\citep{Fey/Lenssen/2019}. 
The code is made available on GitHub\footnote{\url{https://github.com/B-Deprez/AML_Network}}. 

\subsection{Data}
\label{subsec:data}
The methods are compared on two data sets; the Elliptic~\citep{Elliptic, weber2019anti} and IBM money laundering~\citep{altman2024realistic, egressy2024provably} data sets. 
These cover the two main lines of research in the literature. 
The Elliptic data set concerns real-world crypto-transactions, while the IBM data set deals with simulated transactions in fiat currencies. 

\subsubsection{Elliptic.}
The Elliptic data set~\citep{Elliptic, weber2019anti} contains Bitcoin transactions for 49 time steps, each being around two weeks long.
The network has 203~769 nodes and 234~355 edges. The nodes correspond to transactions, and an edge represents that the output of one transaction is the input of the next. 
Additionally, the data set contains 166 pre-calculated features. 
These are split into 94 transaction-specific, i.e., local features and 72 aggregated features, summarizing the local features of a node's neighbours. 

All features are numerical and have already been standardised. 
The local features will be used for all methods. 
Aggregated features are not included for the GNNs, since these methods aggregate the neighbourhood information, including the node features. 
This was also addressed by~\citet{weber2019anti}, stating that graph neural networks are better to address the heterogeneity of the neighbourhoods than the aggregated features. 

Around $33\%$ of transactions are labelled, 4~545 of which as illicit and 42~019 as licit. 
Hence, the classification problem is imbalanced with $2\%$ of all nodes labelled \textit{illicit}. 
We note that an illicit label does not automatically mean that the transaction is used to launder money. 
However, criminals will try to obscure the source of these illicit funds, so we believe that it is still informative. 
Section~\ref{subsec:fullscope} also illustrated that this data set is widely used in AML research. 

We use the Elliptic data set included in the Pytorch Geometric library~\citep{Elliptic_torch,Fey/Lenssen/2019}. 
We split the periods into a train (period 1-30), validation (period 31-40) and test set (period 41-49). 
Only nodes having a label are used for performance evaluation. 
Transactions with label 2 (unknown) are included for the construction of the network, but they do not contribute to training or loss calculation~\citep{kipf2017semisupervised, pmlr-v48-yanga16, weber2019anti}. 

\subsubsection{IBM Money Laundering.}
The second data set is the HI-Small data set, taken from Kaggle, consisting of simulated transactions in a virtual world between individuals, companies and banks~\citep{altman2024realistic, egressy2024provably}. 
A fraction of these transactions are labelled as money laundering. 
Given that the labels are at transaction level, the network is constructed with the transactions as nodes. 
Edges are added from transaction $v_i$ to $v_j$ if the receiver of transaction $v_i$ is the sender of transaction $v_j$, transaction $v_j$ happened before $v_i$, and the difference in time is smaller than $\Delta t$, i.e., $\Delta t \geq t_j-t_i \geq 0$. 
This is inspired by the work by~\citet{tariq2023topologyagnosticdetectiontemporalmoney}, and allows to capture the money flow. 
We take a 60-20-20 train-validation-test split based on transaction time.

Since this results in a massive data set and hyperparameter tuning for DeepWalk and node2vec would take too long, we only take the last 500~000 transactions, and set $\Delta t$ to four hours. 
This results in a network with 500~000 nodes and 1~278~952 edges. 
Just 1356 transactions, i.e., $0.27\%$, are labelled as money laundering.

Each transaction has a couple of features, of which we use the following: `Amount Received', `Receiving Currency', `Amount Paid', `Payment Currency', `Payment Format'. 
Next to that, we extract the day of the week, hour and minute out of a transaction's date stamp. 
The categorical features, i.e., `Receiving Currency', `Payment Currency', `Payment Format', are transformed using one-hot-encoding. 
In the remainder of this paper, we denote this data set as IBM-AML.

\begin{figure}[h!]
			\centering
			{\includegraphics[width = 0.4\textwidth]{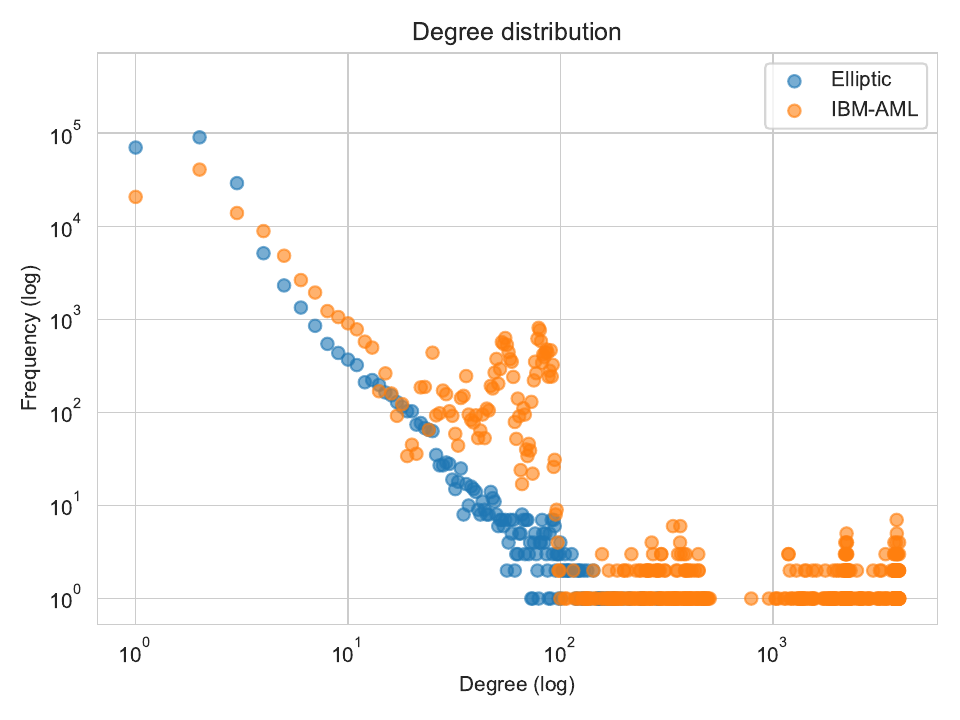}}
			\caption{Comparison of the degree distributions of the Elliptic and IBM-AML data sets.}
            \label{fig:degreeDist}
   {}
\end{figure}

\begin{figure}[h!]
			\centering
			{\includegraphics[width = 0.4\textwidth]{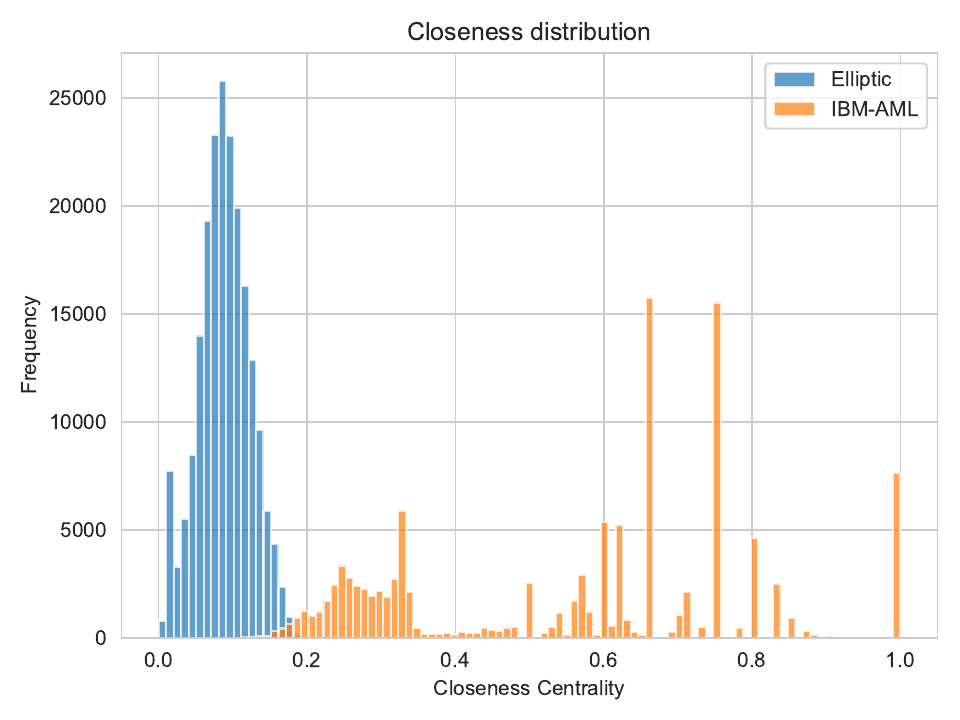}}
			\caption{Comparison of the distributions of the closeness centrality of the Elliptic and IBM-AML data sets.} 
            \label{fig:closenessDist}
   {}
\end{figure}

An analysis of the topology of both networks is given in Table~\ref{tab:networkfeatures}, based on well-known network features. 
IBM-AML seems to be much more connected, compared to the Elliptic data set, having higher average degree and lower average path length. 
This is corroborated by the degree distribution and the distributions of the closeness centrality, in Figure~\ref{fig:degreeDist} and Figure~\ref{fig:closenessDist}, respectively. 
The Elliptic data set more closely presents a scale-free distribution, while the IBM-AML data set deviates from this with fewer nodes with low degree and more nodes with higher degree (hubs). 
The hubs resulted in a much lower average path length, and higher closeness centrality values.

\begin{table}[t]
    \centering
    \caption{Network features for the Elliptic and IBM-AML data set.}
    \label{tab:networkfeatures}
    {
    \begin{tabular}{l|cccccc}
    \toprule
\bf Network & \bf Nodes & \bf Edges &\bf Percentage Fraud & \bf Avg. Degree & \bf Avg. Path Length &\bf Clustering\\ 
\midrule
Elliptic & 203,769 & 234,355 & $2\%$ & 2.3 & 15.96 & 0.0117 \\
IBM-AML & 500,000 & 1,278,952 & $0.27\%$ & 5.1 & 3.78 & 0.0000 \\
    \bottomrule
    \end{tabular}}
{}
\end{table}

\subsection{Hyperparameter Tuning}
\label{subsec:hyperparam}
The hyperparameters are tuned to maximises the AUC-PR on the validation set.
However, it is not feasible to tune these using a grid search. 
As the methods often have many hyperparameters, the number of combinations increases exponentially. 
Additionally, transaction networks are very large, which results in a relative long training time for each individual combination of hyperparameters. 
Therefore, we tune the hyperparameters efficiently via Optuna~\citep{optuna_2019}, whereby we give a range of possible values to select from. 
We use optuna's default \texttt{TPESampler} sampler. 

The ranges are summarised in Table~\ref{tab:hyperparam sup} and~\ref{tab:hyperparam unsup}, for supervised and unsupervised learning, respectively. 
We add superscript S and U to indicate that tuning is done for supervised and unsupervised learning, respectively. 
A model only based on the intrinsic features (IF) is included. 
The number of trials is set to 50 for DeepWalk and node2vec, and 100 for the GNNs.

Since the IBM data set is much larger, scalability in both time and memory are very important. 
Especially DeepWalk and node2vec take much time to run. 
The higher average degree also makes transition probability calculations more expensive. 
That is why the number of epochs for calculating the embedding are capped at 100. 
The attention mechanism of GAT is conditional on the query node, meaning that attention is learned for each node pair. 
To be able to train the model in memory, we limit the attention head to a maximum of two.

\begin{table}[t]
\centering
\caption{The overview of the hyperparameter values for the supervised learning models. The tuned hyperparameter value is added between parentheses next to the name of the model.} 
\label{tab:hyperparam sup}
{\begin{tabular}{L{4cm}|L{2.7cm}L{9cm}}
\toprule
\textbf{Hyperparameter} & \textbf{Tuning Range} & \textbf{Model (hyperparameter value Elliptic $\mid$ IBM )} \\ \midrule
$\alpha$: random jump parameter & $[0.1, 0.9]$ & PageRank$^{\text{S}}$~($0.593\mid 0.646$) \\
Number of walks per node & $[1, 3]\cap \mathbb{N}$ & DeepWalk$^{\text{S}}$~($2\mid2$), node2vec$^{\text{S}}$~($1\mid3$) \\
Walk length & $[3, 10]\cap \mathbb{N}$ & DeepWalk$^{\text{S}}$~($3\mid7$), node2vec$^{\text{S}}$~($9\mid9$) \\
Word2vec context window size & $[2, 10]\cap \mathbb{N}$ & DeepWalk$^{\text{S}}$~($2\mid 2$), node2vec$^{\text{S}}$~($5\mid9$) \\
Latent/Embedding dimension & $[2, 64]\cap \mathbb{N}$ & DeepWalk$^{\text{S}}$~($5\mid52$), node2vec$^{\text{S}}$~($47\mid9$)\\
Latent/Embedding dimension & $[32, 128]\cap \mathbb{N}$ & GCN~($108\mid 97$), GraphSAGE~($48\mid53$), GAT~($99\mid 51$), GIN~($74\mid47$)\\
$p$: return parameter & $[0.5, 2]$ & node2vec$^{\text{S}}$~($1.17\mid0.537$)\\
$q$: in-out parameter & $[0.5, 2]$ & node2vec$^{\text{S}}$~($1.60\mid1.394$)\\
Number of negative samples & $[1, 5] \cap\mathbb{N}$ & DeepWalk$^{\text{S}}$~($1\mid1$), node2vec$^{\text{S}}$~($1\mid4$)\\
GNN hidden dimensions & $[64, 256]\cap\mathbb{N}$ & GCN~($\text{NA}\mid 88$), GraphSAGE~($\text{NA}\mid 175$), GAT~($\text{NA}\mid \text{NA}$), GIN~($\text{NA}\mid \text{NA}$) \\
GNN layers & $[1, 3]\cap\mathbb{N}$ & GCN~($1\mid2$), GraphSAGE~($1\mid 2$), GAT~($1\mid1$), GIN~($1\mid1$) \\
Learning rate & $[0.01, 0.1]$ & IF$^{\text{S}}$~($0.0163\mid 0.0770$), Manual$^{\text{S}}$~($0.0166\mid0.0401$), DeepWalk$^{\text{S}}$~($0.0554\mid 0.0526$), node2vec$^{\text{S}}$~($0.0159\mid0.0708$), GCN~($0.0225\mid0.0514$), GraphSAGE~($0.0248\mid0.0660$), GAT~($0.0421\mid0.0313$), GIN~($0.0345\mid0.0166$) \\
Aggregator & \{min, mean, max\} & GraphSAGE~(mean $\mid$ max) \\
Number of attention heads & $[1, 5]\cap\mathbb{N} \mid [1, 2]\cap\mathbb{N}$ & GAT~($4\mid2$) \\
Dropout rate & $[0, 0.5]$ & GCN~($0.288\mid 0.163$), GraphSAGE~($0.350\mid0.244$), GAT~($0.186\mid 0.188$), GIN~($0.271\mid 0.273$)\\
Number of layers decoder & $[1,3]\cap\mathbb{N}$ & IF$^{\text{S}}$~($1\mid 2$), Manual$^{\text{S}}$~($1\mid2$) \\
Hidden dimension decoder & $[5,20]\cap\mathbb{N}$ & IF$^{\text{S}}$~($5 \mid 14$), Manual$^{\text{S}}$~($6 \mid17$) \\
Number of epochs decoder & $[5,500]\cap\mathbb{N}$ & IF$^{\text{S}}$~($497 \mid 450$ ) \\
Number of epochs decoder & $[5,100]\cap\mathbb{N}$ & Manual$^{\text{S}}$~($64\mid 59$), DeepWalk$^{\text{S}}$~($80\mid75$), node2vec$^{\text{S}}$~($93\mid75$) \\
Number of epochs & $[5,500]\cap\mathbb{N}\mid [5,100]\cap\mathbb{N}$ & DeepWalk$^{\text{S}}$~($176\mid75$), node2vec$^{\text{S}}$~($222\mid84$), GCN~($483\mid 325$), GraphSAGE~($498\mid399$), GAT~($280\mid341$), GIN~($218\mid 89$) \\
\bottomrule
\end{tabular}}
{}
\end{table}

\begin{table}[t]
\centering
\caption{The overview of the hyperparameter values for the unsupervised learning models. The tuned hyperparameter value is added between parentheses next to the name of the model.}
\label{tab:hyperparam unsup}
{\begin{tabular}{L{4cm}|L{2.7cm}L{9cm}}
\toprule
\textbf{Hyperparameter} & \textbf{Tuning Range} & \textbf{Model (hyperparameter value Elliptic $\mid$ IBM )} \\ \midrule
$\alpha$: random jump parameter & $[0.1, 0.9]$ & PageRank$^{\text{U}}$~($0.187\mid0.117$) \\
Number of walks per node & $[1, 3]\cap \mathbb{N}$ & DeepWalk$^{\text{U}}$~($3\mid 2$), node2vec$^{\text{U}}$~($3\mid2$) \\
Walk length & $[3, 10]\cap \mathbb{N}$ & DeepWalk$^{\text{U}}$~($6\mid7$), node2vec$^{\text{U}}$~($4\mid6$) \\
Word2vec context window size & $[2, 10]\cap \mathbb{N}$ & DeepWalk$^{\text{U}}$~($4\mid 6$), node2vec$^{\text{U}}$~($2\mid 4$) \\
Latent/embedding dimension & $[2, 64]\cap \mathbb{N}$ & DeepWalk$^{\text{U}}$~($64\mid6$), node2vec$^{\text{U}}$~($63\mid 47$)\\
$p$: return parameter & $[0.5, 2]$ & node2vec$^{\text{U}}$~($0.78\mid 1.52$)\\
$q$: in-out parameter & $[0.5, 2]$ & node2vec$^{\text{U}}$~($1.58\mid1.60$)\\
Number of negative samples & $[1, 5] \cap\mathbb{N}$ & DeepWalk$^{\text{U}}$~($4\mid5$), node2vec$^{\text{U}}$~($4\mid1$)\\
Learning rate & $[0.01, 0.1]$ & DeepWalk$^{\text{U}}$~($0.0948\mid0.0541$), node2vec$^{\text{U}}$~($0.0414\mid0.062$) \\
Number of epochs & $[5,500]\cap\mathbb{N}\mid [5,100]\cap\mathbb{N}$ & DeepWalk$^{\text{U}}$~($17\mid46$), node2vec$^{\text{U}}$~($59\mid50$) \\
Number of estimators & $[50,200]\cap \mathbb{N}$ & IF$^{\text{U}}$~($60\mid119$), Manual$^{\text{U}}$~($70\mid66$), DeepWalk$^{\text{U}}$~($69\mid169$), node2vec$^{\text{U}}$~($60 \mid 100$)\\
Max. number of samples & $[0.1, 1]$ & IF$^{\text{U}}$~($0.484\mid0.111$), Manual$^{\text{U}}$~($0.333\mid0.559$), DeepWalk$^{\text{U}}$~($0.533\mid0.386$), node2vec$^{\text{U}}$~($0.824\mid 0.354$) \\
Max. number of features $(\%)$ & $\{10*n\mid n\in \mathbb{N}_{10}\}$ & IF$^{\text{U}}$~($0.1\mid0.9$), Manual$^{\text{U}}$~($0.8\mid0.3$), DeepWalk$^{\text{U}}$~($0.8\mid0.2$), node2vec$^{\text{U}}$~($0.8\mid 0.4$) \\
Bootstrap & $\{ \text{True, False} \}$ & IF$^{\text{U}}$~($\text{True}\mid\text{False}$), Manual$^{\text{U}}$~($\text{True}\mid\text{True}$), DeepWalk$^{\text{U}}$~($\text{True}\mid\text{False}$), node2vec$^{\text{U}}$~($\text{True}\mid\text{True}$) \\
\bottomrule
\end{tabular}}
{}
\end{table}

All decoders used for classification have a fixed architecture (Section~\ref{subsec:model}). 
For the GNNs, the training of the decoder weights happens simultaneously with the other weights during training. 
For the manual and shallow representation features, the decoder/classifier needs to be trained afterwards. 
The only hyperparameter tuned is the number of epochs. 

The tuned hyperparameters are presented in parentheses in Table~\ref{tab:hyperparam sup} and Table~\ref{tab:hyperparam unsup}.

\subsection{Performance Metrics}
\label{subsec:metrics}
We tackle money laundering as a binary classification problem. 
We adopt the most popular performance metrics from the literature, namely the precision, recall and F1-score (Section~\ref{subsec:fullscope}). 
These require a classification threshold. 
Since investigation resources are limited~\citep{van2023catchm}, the thresholds will be set relatively high, e.g., by classifying the top $0.1\%$, $1\%$ or $10\%$ of scores as money laundering. 
This is supplemented by a threshold equal to the relative prevalence of money laundering in the data sets.

The selection of these thresholds is ad-hoc, and can influence the conclusions. 
Therefore, we also include threshold-independent metrics. 
The most popular one for binary classification is the area under the ROC curve (AUC-ROC). 
The ROC curve plots the true positive rate against the false positive rate.
However, in highly imbalanced data sets with many negative samples, the false positive rate can appear low even when the absolute number of false positives is high, making it less sensitive to performance issues in detecting the minority class~\citep{davis2006relationship,saito2015precision}. 
The area under the precision-recall curve (AUC-PR) is better suited to compare these models~\citep{davis2006relationship, ozenne2015precision, saito2015precision}. 
This also adds to the literature, since we showed in Section~\ref{subsec:fullscope} that few papers base their conclusions on the AUC-PR.

To test model stability, the models are trained 10 times on a randomised version of the training set.
The manual network features and shallow representation embeddings result in classic tabular features (cf. Figure~\ref{fig:embedding illustration}). 
For these, we apply a bootstrap method, where a new train set is constructed of the same size as the original, where the observations are sampled with replacement. 
The deep representation methods apply train masks. 
We initialise new train masks by randomly selecting half the train masks from the original ones.

\section{Results and Discussion}
\label{sec:res&disc}
\subsection{Supervised Learning}
\label{subsec:res&disc supervised}
The results for the threshold-independent metrics, i.e., AUC-ROC and AUC-PR, on the test set are shown in Table~\ref{tab:resTestTI}, including the standard deviation calculated as described in Section~\ref{subsec:metrics}. 
Additionally, the threshold-dependent results, i.e., precision, recall and F1-score, are given in Table~\ref{tab:resTestTD_1_Elliptic} and Table~\ref{tab:resTestTD_1_IBM} for the top-$1\%$ results. 
To analyse the sensitivity to the threshold, we report the results on the other three thresholds in Appendix~C. 

\begin{table}
    \centering
    \caption{Threshold-independent metrics: AUC-ROC and AUC-PR values over the different methods for the two data sets, based on the test set. The standard deviation is also reported.} 
    \label{tab:resTestTI}
    {\begin{tabular}{l|cc|cc}
    \toprule
    \bf Supervised& \multicolumn{2}{c}{\bf Elliptic}
    & \multicolumn{2}{c}{\bf IBM-AML}\\
\bf Methods & \bf AUC-ROC & \bf AUC-PR & \bf AUC-ROC & \bf AUC-PR   \\ 
\midrule
    Intrinsic features & $0.8607 \pm 0.0079$ & $0.5730 \pm 0.0958$ & $0.7560 \pm 0.0075$ & $0.0186 \pm 0.0008$ \\
    Egonet features & $0.8648 \pm 0.0090$ & $0.5956 \pm 0.0328$ & $\boldsymbol{0.7579 \pm 0.0033}$ & $\boldsymbol{0.0187 \pm 0.0004}$ \\
    DeepWalk & $0.8483 \pm 0.0083$ & $0.5823 \pm 0.0265$ & $0.7558 \pm 0.0033$ & $0.0185 \pm 0.0004$ \\ 
    Node2vec & $0.8497 \pm 0.0044$ & $0.5938 \pm 0.0251$ & $0.7551 \pm 0.0031$ & $0.0184 \pm 0.0004$\\ 
    GCN &   $0.8465 \pm 0.0110$ & $0.5948 \pm 0.0164$ & $0.5657 \pm 0.0904$ & $0.0113 \pm 0.0029$ \\
    GraphSAGE & $\boldsymbol{0.8712 \pm 0.0122}$ & $\boldsymbol{0.6392 \pm 0.0336}$ & $0.6068 \pm 0.0793$ & $0.0120 \pm 0.0028$ \\
    GAT &  $0.8579 \pm 0.0161$ & $0.6376 \pm 0.0270$  & $0.4203 \pm 0.1889$ & $0.0103 \pm 0.0041$ \\
    GIN & $0.8213 \pm 0.0203$ & $0.5079 \pm 0.0593$  & $0.5028 \pm 0.2533$ & $0.0119 \pm 0.0065$ \\
    \bottomrule
    \end{tabular}}
    {}
\end{table}

\begin{table}
    \centering
    \caption{Threshold-dependent metrics: Precision, recall and F1-score values for the top $1\%$ scores over the different methods for the Elliptic data set, based on the test set. The standard deviation is also reported.}
    \label{tab:resTestTD_1_Elliptic}
    {\begin{tabular}{l|ccc}
    \toprule
    & \multicolumn{3}{c}{\bf Elliptic Supervised}  \\
\bf Methods & \bf Precision & \bf Recall  & \bf F1-score   \\ 
\midrule
    Intrinsic features & $0.8570 \pm 0.2706$ & $0.1840 \pm 0.1173$ & $0.2932 \pm 0.1451$  \\
    Egonet features & $0.9339 \pm 0.0463$ & $0.1645 \pm 0.0082$ & $0.2797 \pm 0.0139$  \\
    DeepWalk & $0.9116 \pm 0.0464$ & $\boldsymbol{0.1881 \pm 0.0895}$ & $\boldsymbol{0.3049 \pm 0.1054}$ \\ 
    Node2vec & $0.9321 \pm 0.0465$ & $0.1642 \pm 0.0082$ & $0.2791 \pm 0.0139$  \\
    GCN & $0.9911 \pm 0.0094$ & $0.1728 \pm 0.0047$ & $0.2942 \pm 0.0070$ \\
    GraphSAGE & $0.9822 \pm 0.0188$ & $0.1721 \pm 0.0077$ & $0.2929 \pm 0.0113$ \\
    GAT & $\boldsymbol{0.9964 \pm 0.0075}$ & $0.1753 \pm 0.0067$ & $0.2981 \pm 0.0099$\\
    GIN & $0.9286 \pm 0.0709$ & $0.1633 \pm 0.0129$ & $0.2777 \pm 0.0214$\\
    \bottomrule
    \end{tabular}}
    {}
\end{table}

\begin{table}
    \centering
    \caption{Threshold-dependent metrics: Precision, recall and F1-score values for the top $1\%$ scores over the different methods for the IBM-AML data set, based on the test set. The standard deviation is also reported.}
    \label{tab:resTestTD_1_IBM}
    {\begin{tabular}{l|ccc}
    \toprule
    & \multicolumn{3}{c}{\bf IBM-AML Supervised}  \\
\bf Methods & \bf Precision & \bf Recall  & \bf F1-score   \\ 
\midrule
    Intrinsic features & $0.0190 \pm 0.0040$ & $0.0210 \pm 0.0044$ & $0.0200 \pm 0.0042$ \\
    Egonet features & $\boldsymbol{0.0192 \pm 0.0034}$ & $0.0213 \pm 0.0038$ & $0.0202 \pm 0.0036$ \\
    DeepWalk & $0.0174 \pm 0.0053$ & $0.0193 \pm 0.0059$ & $0.0183 \pm 0.0056$ \\ 
    Node2vec & $0.0181 \pm 0.0047$ & $0.0200 \pm 0.0052$ & $0.0190 \pm 0.0049$ \\
    GCN &  $0.0100 \pm 0.0045$ & $0.8807 \pm 0.3119$ & $0.0197 \pm 0.0089$ \\
    GraphSAGE & $0.0119 \pm 0.0027$ & $\boldsymbol{0.9742 \pm 0.0298}$ & $\boldsymbol{0.0236 \pm 0.0052}$ \\
    GAT & $0.0065 \pm 0.0070$ & $0.1805 \pm 0.3695$ & $0.0104 \pm 0.0139$ \\
    GIN & $0.0085 \pm 0.0087$ & $0.0166 \pm 0.0087$ & $0.0092 \pm 0.0089$ \\
    \bottomrule
    \end{tabular}}
    {}
\end{table}

For both data sets, the models with intrinsic features give already good result. 
Adding egonet features is beneficial for both data sets. 
We do, however, see a drop in performance for DeepWalk and node2vec.
It is possible that some of the feature values, i.e., the coordinates in the latent space, are correlated or noisy. 
Neural networks, used here as the decoder for classification, are known to suffer under correlated and noisy features~\citep{grinsztajn2022tree}.
GIN also performs poorly on both data sets. 
One explanation is that the neural network in the aggregation step is not optimal, and should be tuned further. 
Another explanation is that the inclusion of neural network aggregation makes GIN more sensitive to the high class imbalance. 
Hence, GIN seems less suited for anti-money laundering. 
It should be noted that good performance is observed for the precision indicating that GIN reduces false positives in the top-percentiles of predictions. 

When looking at the Elliptic data set, we see that also the GNNs appear to underperform, except for GraphSAGE. 
Additionally, GAT has good AUC-PR performance. 
We see in Tables~\ref{tab:resTestTD_1_Elliptic},~\ref{tab:resTestTD_01_Elliptic},~\ref{tab:resTestTD_10_Elliptic} and~\ref{tab:resTestTD_p_Elliptic} that for the top percentages the GNNs have remarkably higher precision, but lower recall. 
Although fewer cases are detected, those with highest predicted money laundering propensity contain far fewer false positives. 
This is highly important when adopting these models in practice.

The results on the IBM data set in Table~\ref{tab:resTestTI} show a different picture. 
Only the egonet features improve performance. 
The threshold-dependent metrics show a similar picture.

One explanation for the poor performance of GNNs might be related to the network structure (Section~\ref{subsec:data}), with very short average paths and many high-degree nodes. 
This results in strong over-smoothing~\citep{Li_Han_Wu_2018}, resulting in very similar embeddings for the different nodes with insufficient discriminatory power.

Moreover, a deeper analysis reveals that the training loss, when training on the train and validation set, is highly unstable.
Depending on the epoch and specific weight initialization, the results vary widely, with the models sometimes just predicting the majority class. 
For this specific data set, GNNs seem to face instability due to the extreme class imbalance.

The supervised learning experiments on both data sets demonstrate that the network structure provides additional insights for anti-money laundering. 
The observed performance improvements are limited, but even a small relative increase in performance has material implications for the business. 
The efficiency of AML practices increases significantly since investigators have more high quality predictions, and spend less time on the investigation of false positives. 

However, depending on the network topology and the extent of the class imbalance, GNNs might be too unstable. 
Additional experiments are needed to quantify the sensitivity of GNNs to the imbalance, but this is outside the scope of this research.

\subsection{Unsupervised Learning}
\label{subsec:res&disc unsupervised}
The results of the unsupervised methods using isolation forests are given in Tables~\ref{tab:resTestTI_unsup}. 
None of the results on the Elliptic data set are particularly good, with AUC-ROC values below 0.5. 
For completeness, we also include the threshold-dependent metric in Table~\ref{tab:resTestTDunsup_1_Elliptic}, where most models fail to detect any money laundering cases. 

These results are in line with a previous study performed by \citet{10.1145/3383455.3422549} who tested anomaly detection methods on the Elliptic data set. 
The authors did not include any additional network analytics methods and only reported the F1 score. 
However, they found that across the methods, the anomaly detection results were much worse than for supervised learning, with reported F1-scores for the isolation forest equal to 0.

\begin{figure}[h!]
			\centering
			{\includegraphics[width = 0.6\textwidth]{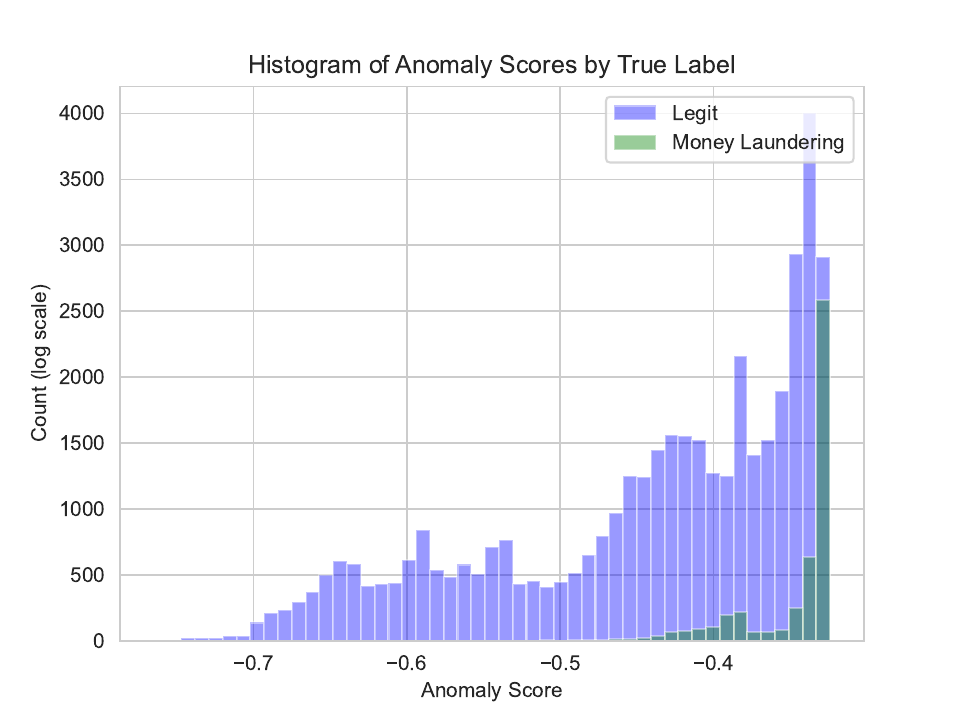}}
			\caption{The anomaly scores for the legitimate and money laundering cases. More negative is more anomalous.}
            \label{fig:anomalyIF}
   {}
\end{figure}

We perform additional analyses on these results.
Figure~\ref{fig:anomalyIF} shows the anomaly scores for the intrinsic features for the Elliptic data set. 
Here, a very negative score signifies anomalies. 
All money laundering cases have scores near zero, within the bulk of the distribution, indicating that the isolation forest does not recognize these cases as anomalies.

Conversely, the results for IBM-AML are more in line with expectations. 
Network information captured by DeepWalk, results in the best performing model here. 
Comparing these results to those in Table~\ref{tab:resTestTI}, we see that DeepWalk even outperforms the supervised learning methods in terms of AUC-PR.

\begin{table}
    \centering
    \caption{Threshold-independent metrics: AUC-ROC and AUC-PR values over the different methods for the two data set, based on the test set. The standard deviation is also reported.}
    \label{tab:resTestTI_unsup}
    {\begin{tabular}{l|cc|cc}
    \toprule
    \bf Unsupervised& \multicolumn{2}{c}{\bf Elliptic}
    & \multicolumn{2}{c}{\bf IBM-AML}\\
\bf Methods & \bf AUC-ROC & \bf AUC-PR & \bf AUC-ROC & \bf AUC-PR   \\ 
\midrule
    Intrinsic features & $0.134 \pm 0.007$ & $0.058 \pm 0.001$ &$0.710 \pm 0.015$ & $0.017 \pm 0.003$ \\
    Egonet features & $0.217 \pm 0.010$ & $0.013 \pm 0.000$ & $0.700 \pm 0.033$ & $0.011 \pm 0.005$ \\
    DeepWalk &$0.182 \pm 0.008$ & $0.056 \pm 0.001$ & $\boldsymbol{0.744 \pm 0.023}$ & $\boldsymbol{0.024 \pm 0.008}$\\ 
    Node2vec & $0.190 \pm 0.008$ & $0.056 \pm 0.0005$ & $0.632 \pm 0.031$ & $0.005 \pm 0.001$ \\ 
    \bottomrule
    \end{tabular}}
    {}
\end{table}

\begin{table}
    \centering
    \caption{Threshold-dependent metrics: Precision, recall and F1-score values for the top $1\%$ scores over the different methods for the Elliptic data set, based on the test set. The standard deviation is also reported.} 
    \label{tab:resTestTDunsup_1_Elliptic}
    {\begin{tabular}{l|ccc}
    \toprule
    & \multicolumn{3}{c}{\bf Elliptic Unsupervised}  \\
\bf Methods & \bf Precision & \bf Recall  & \bf F1-score   \\ 
\midrule
    Intrinsic features & $0.000 \pm 0.000$ & $0.000 \pm 0.000$ & $0.000 \pm 0.000$ \\
    Egonet features & $0.000 \pm 0.000$ & $0.000 \pm 0.000$ & $0.000 \pm 0.000$ \\
    DeepWalk & $0.005 \pm 0.005$ & $0.001 \pm 0.001$ & $0.001 \pm 0.001$\\ 
    Node2vec & $0.003 \pm 0.004$ & $0.000 \pm 0.000$ & $0.001 \pm 0.001$ \\
    \bottomrule
    \end{tabular}}
    {}
\end{table}

\begin{table}
    \centering
    \caption{Threshold-dependent metrics: Precision, recall and F1-score values for the top $1\%$ scores over the different methods for the IBM-AML data set, based on the test set. The standard deviation is also reported.}
    \label{tab:resTestTDunsup_1_IBM}
    {\begin{tabular}{l|ccc}
    \toprule
    & \multicolumn{3}{c}{\bf IBM-AML Unupervised}  \\
\bf Methods & \bf Precision & \bf Recall  & \bf F1-score   \\ 
\midrule
    Intrinsic features & $0.042 \pm 0.005$ & $0.156 \pm 0.020$ & $0.067 \pm 0.008$\\
    Egonet features & $0.027 \pm 0.013$ & $0.098 \pm 0.048$ & $0.042 \pm 0.021$ \\
    DeepWalk & $\boldsymbol{0.055 \pm 0.009}$ & $\boldsymbol{0.204 \pm 0.034}$ & $\boldsymbol{0.087 \pm 0.015}$ \\ 
    Node2vec & $0.007 \pm 0.002$ & $0.025 \pm 0.008$ & $0.010 \pm 0.003$ \\
    \bottomrule
    \end{tabular}}
    {}
\end{table}

The most plausible explanation in our opinion lies in the assumptions underlying isolation forest and the data generating process.
Two specific assumptions by~\citet{lui2012isolation}:
\begin{itemize}
    \item Anomalies yield fewer partitions since anomalies occupy regions with lower density;
    \item Instances with distinct attributes are separated earlier in the tree, contributing to shorter paths.
\end{itemize}

Therefore, the nature of the data sets explains the difference in performance. 
The Elliptic data set is a real-life data set, containing criminals that actively attempt to cover their tracks~\cite{baesens2015fraud}. 
They try to mimic the `average behaviour', but no real transaction/person is average on all attributes. 
This is why it seems that classifying the least suspicious transactions as money launderers, we would achieve an AUC-ROC around $80\%$. 
Hence, transactions that appear `too normal' are actually most suspicious. 
In contrast, the IBM-AML data set is synthetic, with fraudulent patterns and behaviours deliberately introduced. 
It is likely that camouflage behaviour is not present in this data set. 

These insight highlights the importance of knowing the exact assumptions underlying the data sets used. 
Synthetic data sets might give overly optimistic results, since complex behaviour inherent to the problem at hand are not captured. 

\subsection{Ablation Study}
To better understand the effect of the network structure on the predictive power of the models, we analyse the performance of the different models without intrinsic features. 
For the models, we take the hyperparameters as determines in Table~\ref{tab:hyperparam sup} and Table~\ref{tab:hyperparam unsup}. 
The GNNs require feature values, so we give each node a dummy feature with value 1~\citep{van2022inductive}. 
Additionally, for the Elliptic data set, we have no control on the summary features, so we discard these as well, although they represent some network structure.

The results for supervised learning are given in Table~\ref{tab:resTestTI_ablation}. 
The AUC values indicate that most methods based purely on the network structure are close to random. 

One notable exception is GIN, which is specifically designed to better distinguish certain graph structures~\citep{xu2019powerful}. 
Graph structure is the only information available in this part of the experiments.
The results for GIN indicate that there are some structural difference in the network for fraudulent and non-fraudulent nodes in the Elliptic data set. These structural difference are complex, since they are not picked up by the egonet features or the random-walk-based ones. 

The other GNN architectures struggle, since they rely heavily on the node features. 
Especially GAT---which performed well with features---leverages node features for the attention calculations. 

For the IBM data set, the differences are less pronounced that for the Elliptic data set. 
The network structure alone might not be enough to uncover money laundering. 
The data generation process of this synthetic data set creates a direct correlation between the money laundering pattern and feature values. 
Therefore, these feature values are more important than the network structure.

\begin{table}
    \centering
    \caption{Threshold-independent metrics: AUC-ROC and AUC-PR values over the different methods for the two data sets, based on the test set. The standard deviation is also reported.}
    \label{tab:resTestTI_ablation}
    {\begin{tabular}{l|cc|cc}
    \toprule
    \bf Supervised& \multicolumn{2}{c}{\bf Elliptic}
    & \multicolumn{2}{c}{\bf IBM-AML}\\
\bf Methods & \bf AUC-ROC & \bf AUC-PR & \bf AUC-ROC & \bf AUC-PR   \\ 
\midrule
    Egonet features & $0.5172 \pm 0.0607$ & $0.0630 \pm 0.0122$ & $0.4758 \pm 0.0187$ & $0.0088 \pm 0.0003$ \\
    DeepWalk & $0.5125 \pm 0.0184$ & $0.0597 \pm 0.0021$ & $0.5019 \pm 0.0143$ & $0.0092 \pm 0.0004$ \\ 
    Node2vec & $0.4454 \pm 0.0118$ & $0.0511 \pm 0.0021$ & $0.4931 \pm 0.0095$ & $0.0091 \pm 0.0003$ \\ 
    GCN & $0.5655 \pm 0.0139$ & $0.0696 \pm 0.0039$ & $0.4985 \pm 0.0183$ & $0.0092 \pm 0.0007$ \\
    GraphSAGE & $0.5000 \pm 0.0000$ & $0.0565 \pm 0.0025$ & $0.5050 \pm 0.0081$ & $\boldsymbol{0.0093 \pm 0.0003}$ \\ 
    GAT & $0.5002 \pm 0.0032$ & $0.0566 \pm 0.0018$ & $0.5014 \pm 0.0035$ & $0.0091 \pm 0.0002$ \\
    GIN & $\boldsymbol{0.6177 \pm 0.0110}$ & $\boldsymbol{0.0739 \pm 0.0049}$ & $\boldsymbol{0.5061 \pm 0.0035}$ & $0.0092 \pm 0.0003$ \\
    \bottomrule
    \end{tabular}}
    {}
\end{table}

The results for the isolation forest in Table~\ref{tab:resTestTI_ablation_unsupervised} show an improvement for the Elliptic data set, compared to the isolation forest with intrinsic features. 
Here, however, the models mostly make random predictions. 
When including the features, a better distinction can be made between money and non-money laundering. 
As mentioned before, the features of the money laundering transactions are so average (not anomalous according to the isolation forest) that they become suspicious again. 
The inverse predictions were in that case very good predictors. 
When only considering the network structure, it is harder to find meaningful anomalies. 
Hence, a higher AUC in this case points to a less useful model, as the model is as good as random.

The results on the IBM data set are in line with what we have seen before. 
The model makes almost random predictions, with deepwalk giving the best AUC-ROC and AUC-PR values.

\begin{table}
    \centering
    \caption{Threshold-independent metrics: AUC-ROC and AUC-PR values over the different methods for the two data sets, based on the test set. The standard deviation is also reported.}
    \label{tab:resTestTI_ablation_unsupervised}
    {\begin{tabular}{l|cc|cc}
    \toprule
    \bf Supervised& \multicolumn{2}{c}{\bf Elliptic}
    & \multicolumn{2}{c}{\bf IBM-AML}\\
\bf Methods & \bf AUC-ROC & \bf AUC-PR & \bf AUC-ROC & \bf AUC-PR   \\ 
\midrule
    Egonet features & $\boldsymbol{0.5254 \pm 0.0120}$ & $0.0226 \pm 0.0008$ & $0.4737 \pm 0.0057$ & $0.0027 \pm 0.0000$ \\
    DeepWalk & $0.4337 \pm 0.0064$ & $\boldsymbol{0.0834 \pm 0.0013}$ & $\boldsymbol{0.5380 \pm 0.0081}$ & $\boldsymbol{0.0031 \pm 0.0001}$ \\ 
    Node2vec & $0.4231 \pm 0.0050$ & $0.0815 \pm 0.0010$ & $0.4910 \pm 0.0092$ & $0.0026 \pm 0.0001$ \\ 
    \bottomrule
    \end{tabular}}
    {}
\end{table}

This ablation study showed that it seems that network structure alone is not enough to come to a performant AML model. 
There is an important interplay between the network and the node features, pointing to a need to combine the two. 
Additionally, it highlighted the strong dependence of GNNs on these node features---except for the GIN in some cases---and on the data generation process. 
The explicit correlation between the money laundering cases and feature values for the IBM data set resulted in sub-optimal results when only considering the network structure. 

\subsection{Model Stability}
\begin{figure}[t]
			\centering
			{\includegraphics[width = \textwidth]{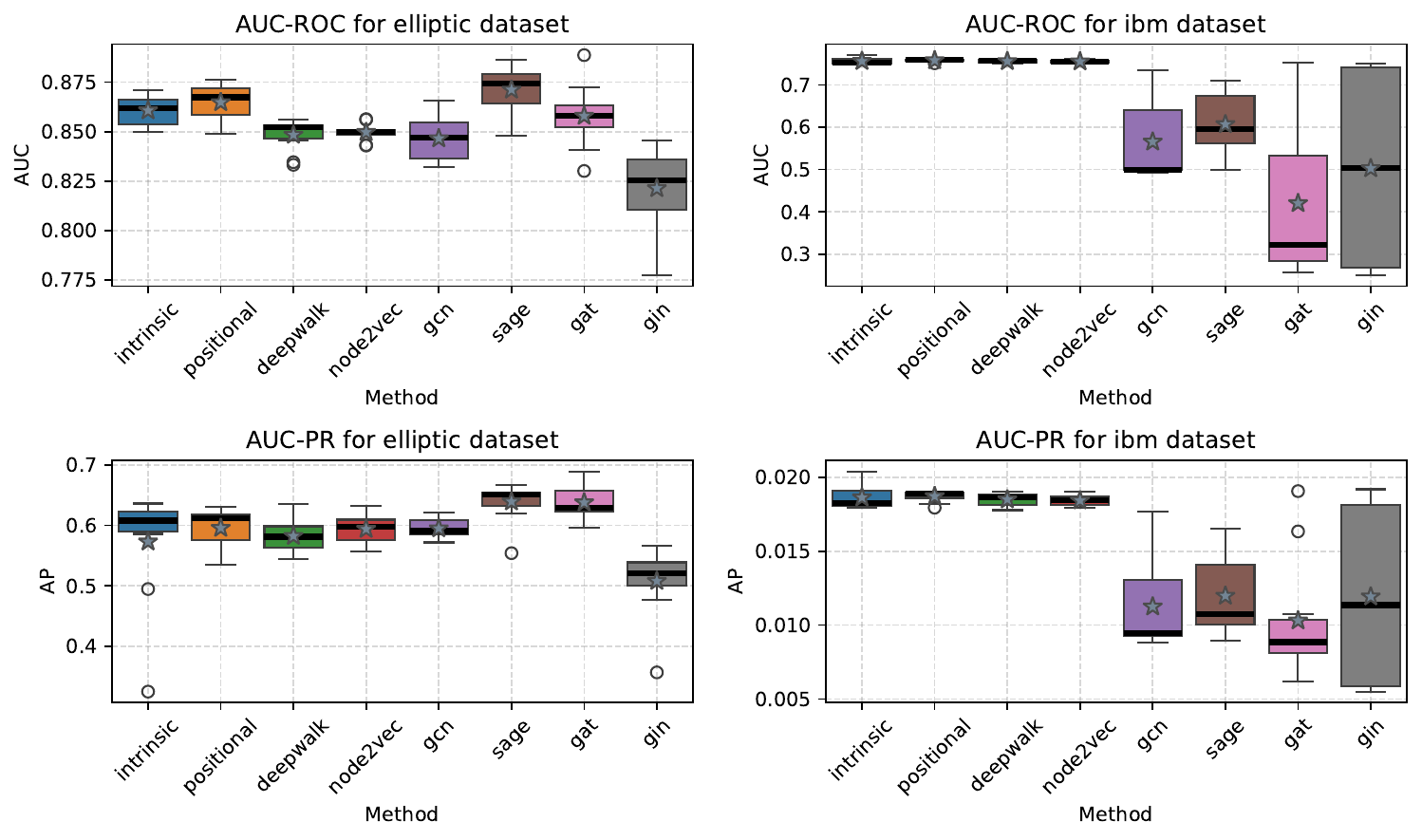}}
			\caption{Box plots of the AUC-ROC and AUC-PR for the ten different training sets.}
            \label{fig:ROC_PR_boxplot}
   {}
		\end{figure}

As a final analysis, we look at the range of the results to assess model stability. 
Figure~\ref{fig:ROC_PR_boxplot} present the box plots of the results for the models on the ten different training sets.

The results are quite stable on the Elliptic data set, although the AUC-ROC shows a bit more variability for the GNNs. 
The values on the IBM data set show strong instability for all GNNs, while the other models are relatively stable. 
The GNNs are probably less stable on the IBM data set due to the larger class imbalance. 

\section{Conclusion}
\label{sec:conclusion}
\subsection{Literature Review}
This paper present a literature review on network analytics for anti-money laundering. 
The 97 selected papers were processed in two steps to achieve (1) a comprehensive understanding of the broader literature and (2) detailed insight into research that has been well received by the community. 
The comprehensive analysis classified all papers according to seven categories, each having different sub-characteristics. 
The detailed analysis was performed on the $10\%$ most-cited papers in scope, which were further categorized using the framework recently proposed by~\citet{bockel2023fraud}. 
Information was given on the different processing steps taken, the challenges faced or addressed in the papers, the experiments performed, and the definition of nodes and edges in the networks. 

A large part of the literature concerns methods that rely heavily on expert-knowledge by applying manual feature engineering and rule-based methods. 
Most of these are based on basic network centrality metrics.

The analysis and comparison of the models is mostly done using threshold-dependent metric, most notably accuracy, precision and recall. 
Given the class imbalance, we strongly advise against the use of accuracy as an evaluation metric for AML models. 
Instead, the AML research should adopt the threshold-independent AUC-PR as a standard evaluation metric, as this is well suited for model comparison in the presence of high class imbalance. 

Our analysis of the literature identified six critical gaps that future research must address:
\begin{enumerate}
    \item \textbf{Benchmark}: There is a lack of comparison of novel methods with other models in the literature. This shows that there is a need for a unified approach to perform test including standard open-source data sets on which to test as well as a consensus on which methods can be seen as baseline models and which as state-of-the-art. 
    \item \textbf{Interpretability}: Few studies have explored tools to explain the outputs of graph neural networks (GNNs), which is essential for regulatory and operational transparency.
    \item \textbf{Unsupervised Methods}: Limited exploration of classic unsupervised methods restricts their potential to detect anomalies, a key requirement as money laundering tactics evolve.
    \item \textbf{Dynamic Network Analysis}: Most approaches rely on static networks, with insufficient research into dynamic methods that could better capture the temporal nature of transactions. 
    \item \textbf{Fraud Specific Methods}: The inherent incompleteness of AML labels, due to undetected cases and resource constraints, highlights the need for specialized PU-learning techniques, while limited resources necessitate actionable, prioritized predictions, which could be achieved through learning-to-rank methods.
    \item \textbf{Adoption State-of-the-Art:} AML is often done using basic network metrics, while new developments in the field are adopted slowly. There is a need for standardising the use of heterogeneous and dynamic networks in AML. Future work should study how this can be achieved by comparing the latest state-of-the-art in continual graph learning, graph contrastive learning, heterogeneous and temporal GNNs, and graph transformers for AML, to name a few.
\end{enumerate}

\subsection{Experimental Evaluation}
To extend upon the existing body of knowledge and to partially address the gaps in the literature, we implemented an extensive experimental evaluation covering manual feature engineering, shallow and deep representation learning, both in a supervised and unsupervised setting. 
We can that network features bring additional predictive power to AML models, when combined with the node features.
However, the GNNs struggled when faced with extreme class imbalance and with a network with many hubs.

The observed performance improvements were limited, but even a small relative increase in performance has material implications for the business. 
The efficiency of AML practices increases significantly since investigators have more high quality predictions, and spend less time on the investigation of false positives. 

In the unsupervised learning experiments, we employed isolation forests and observed notable differences between the real-world Elliptic data set and the synthetic IBM-AML data set. 
Although on synthetic data, the methods performed well, the observations from the Elliptic data seemed to exhibit strong camouflaging tactics. 
The anomaly scores of the latter were so low that investigating the least suspicious transactions for money laundering became the better strategy. 
We conclude from this that care should be taken when testing methods on synthetic data, since it can give overly optimistic results. 

Our experiments on unsupervised learning are, however, limited. 
We only applied an isolation forest, which is a global anomaly detection method. 
Future work should focus on extending this to a larger benchmark, in line with the work of \citet{10.1145/3383455.3422549}, to analyse if the same conclusions hold for other global and local anomaly detection methods. 

The additional ablation study highlighted the need to incorporate both the network and feature information. 
When excluding the node features, the model predictions are as good as random. 
While this approach is common in the literature, we recommend against relying solely on the network structure. 

\subsection{Limitations}
There are limitations to our work that should be addressed in future research.

First, conclusions are drawn based on only two data sets. To generalise our conclusions, additional tests need to be performed on a broader range of data sets. 
This can be done by including more open-source data sets when these become available, and rely on practitioners to use our code to test the methods on their proprietary data and publish the results.

Second, the decoders were always based on neural networks. The application and impact analysis of applying different machine learning methods as down-stream classifiers is another interesting area of study to extend the experiments presented here. 

Third, the hyperparameter tuning process was limited. 
While Optuna provided an efficient method for hyperparameter search, the 50 to 100 tuning rounds used in this study may be insufficient.
The choices in this paper were made to keep everything tractable and feasible given the limited resources available. 
Future work could expand the parameter ranges and number of tuning rounds. 

\section*{Acknowledgments}
This work was supported by the Research Foundation – Flanders (FWO research project 1SHEN24N) and by the BNP Paribas Fortis Chair in Fraud Analytics. 
The resources and services used in this work were provided by the VSC (Flemish Supercomputer Center), funded by the Research Foundation - Flanders (FWO) and the Flemish Government.

\bibliographystyle{plainnat} 
\bibliography{references}

\clearpage

\appendix

\counterwithin{figure}{section}
\counterwithin{table}{section}
\renewcommand\thefigure{\thesection\arabic{figure}}
\renewcommand\thetable{\thesection\arabic{table}}

\section{Analysis of Categories between Crypto and Non-Crypto Papers}
\label{appendix crypto}
This part of the appendix gives the plots in which we compare the difference in nature of the crypto-literature compared non-crypto-related research. 

{
\begin{figure}[t]
			\centering
			{\includegraphics[height = 0.4\textwidth]{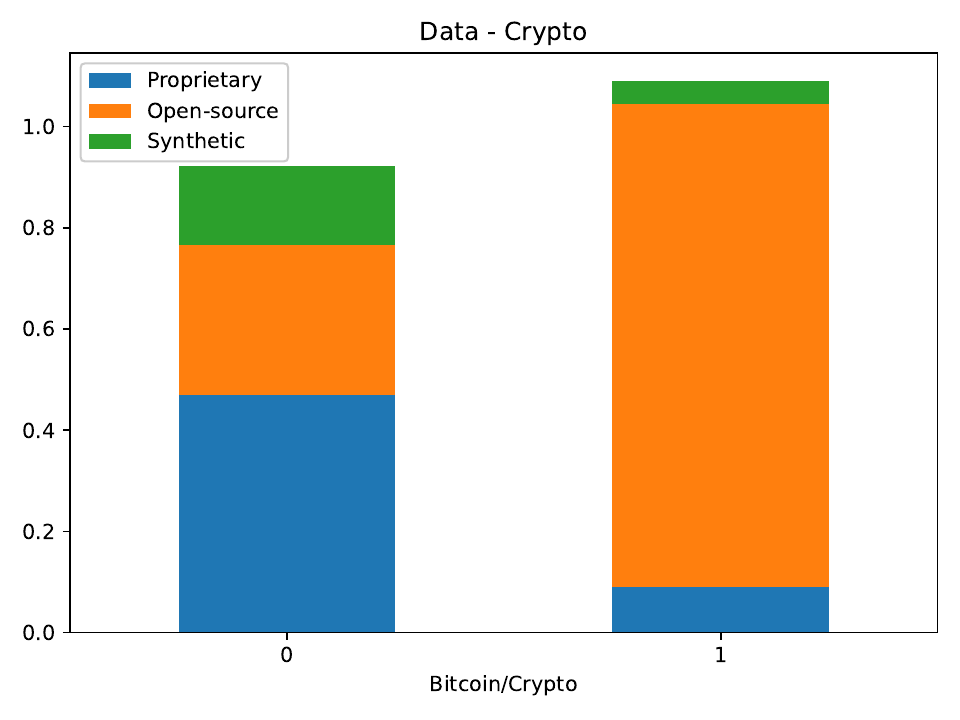}}
			\caption{The data for papers dealing with crypto currencies (1) and those that do not (0).}
            \label{subfig:cryptodata}
   {}
		\end{figure}

\begin{figure}[h!]
		\centering
		{\begin{subfigure}[t]{0.45\textwidth}
			\centering
			\includegraphics[height = 0.7\textwidth]{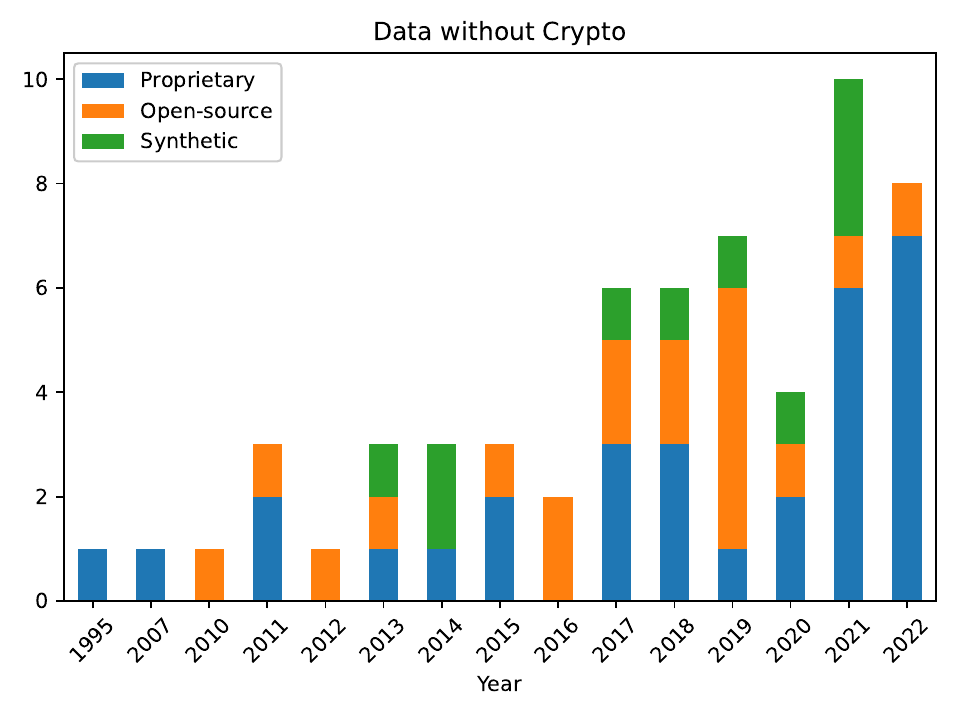}
			\caption{The evolution of the data over the years for papers not covering crypto currencies. }
			\label{subfig:nocryptodistdatayear}
		\end{subfigure} 
        \begin{subfigure}[t]{0.45\textwidth}
			\centering
			\includegraphics[height = 0.7\textwidth]{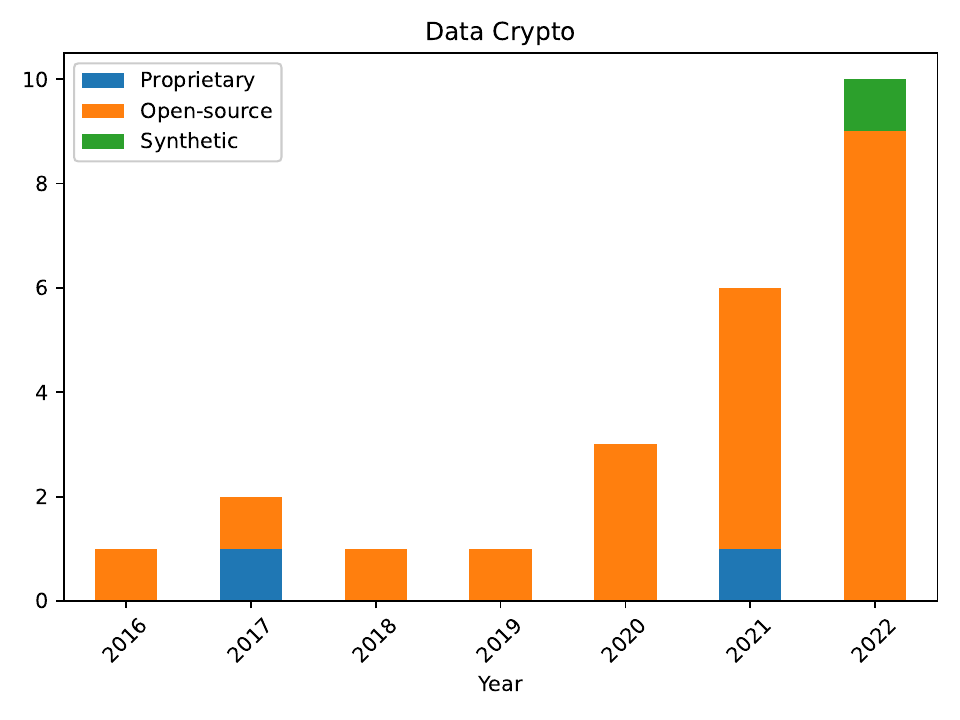}
			\caption{The evolution of the data over the years for papers covering crypto currencies. }
			\label{subfig:cryptodistdatayear}
		\end{subfigure} }
  \caption{Evolution of the data.}
  {}
\end{figure}

\begin{figure}[h!]
			\centering
			{\includegraphics[width = 0.4\textwidth]{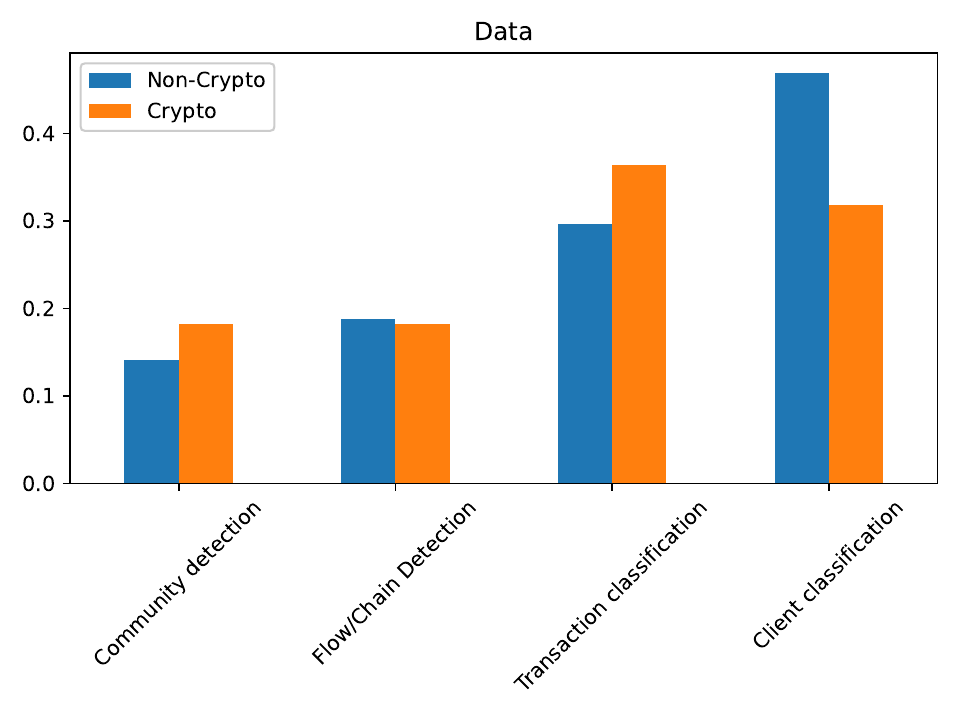}}
			\caption{The distribution of the objective of the papers.}
            \label{fig:cryptodistobjyear}
   {}
\end{figure}

\begin{figure}[h!]
			\centering
			{\includegraphics[width = 0.4\textwidth]{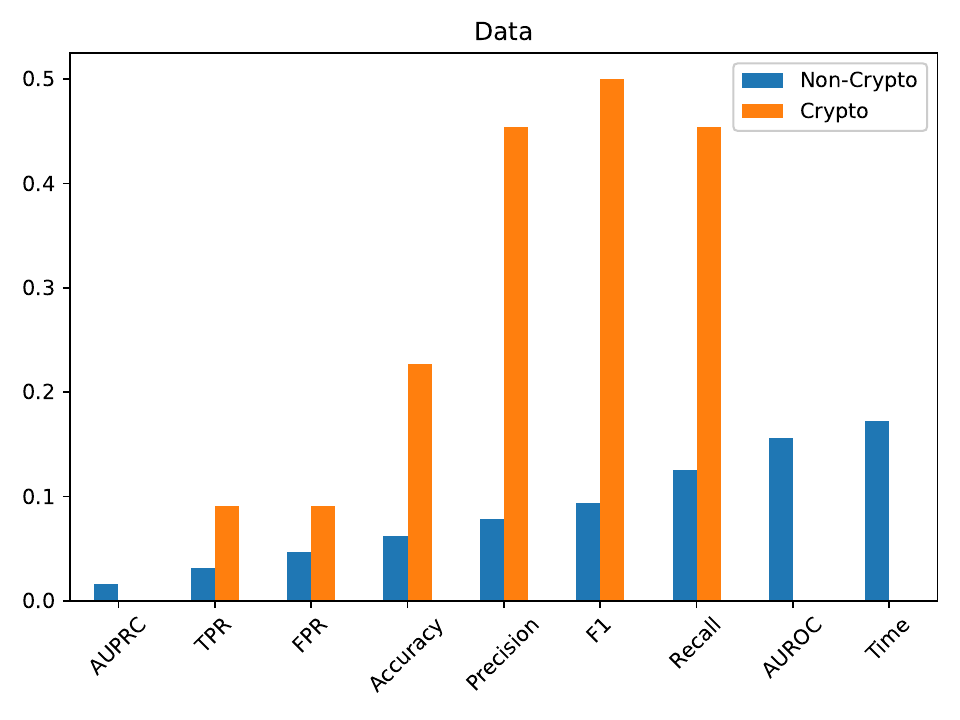}}
			\caption{The distribution of the evaluation metrics used. }
            \label{fig:cryptodistmetryear}
   {}
	\end{figure}
}

\clearpage

\section{The Top-Cited Papers}
\label{appendix top}
In this part of the appendix, we give the figures that summarise the top-cited papers. 

{
\begin{figure}[h!]
\centering
{\includegraphics[width = 0.4\textwidth]{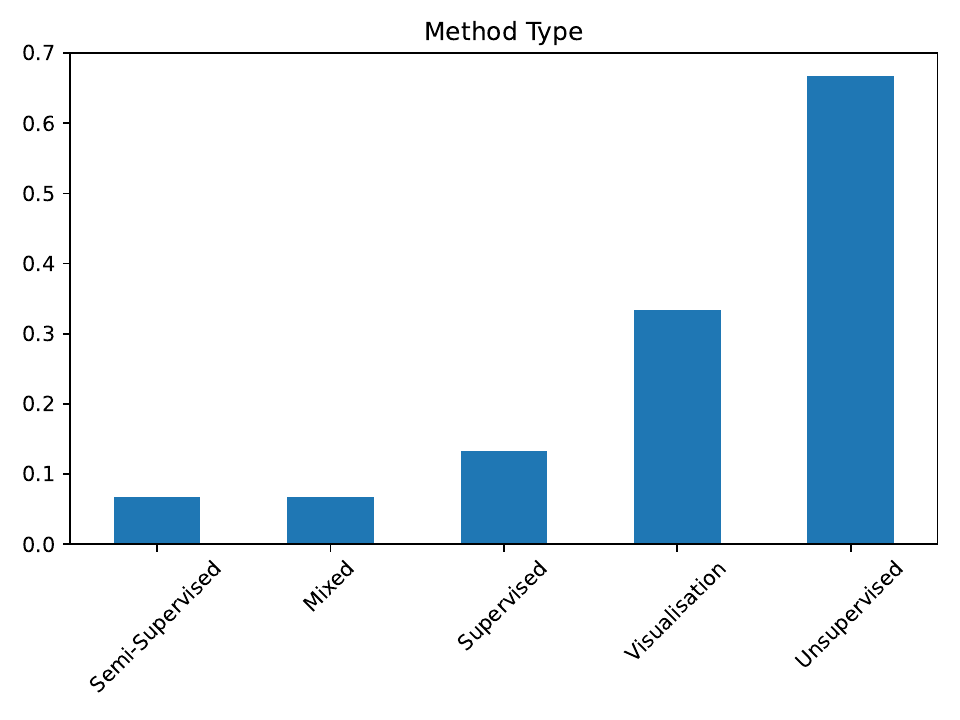}}
\caption{The methods for the top-cited papers.}
\label{fig:toptech}
{}
\end{figure} 

\begin{figure}[h!]
\centering
{\includegraphics[width = 0.4\textwidth]{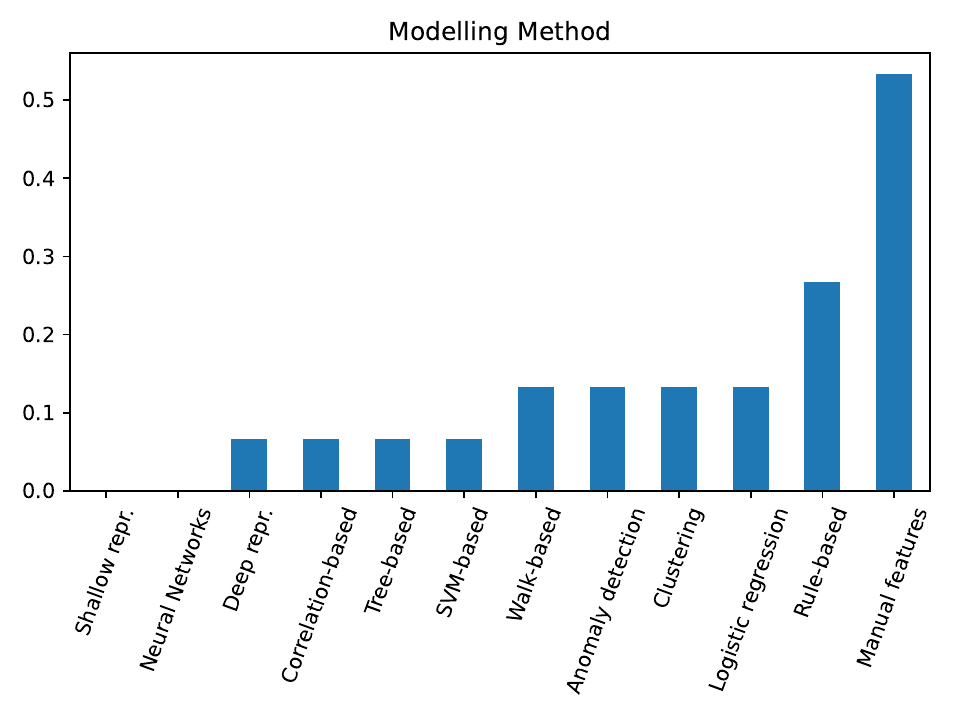}}
\caption{The modelling methods for the top-cited papers.}
\label{fig:topdetail}
{}
\end{figure} 

\begin{figure}[h!]
\centering
{\includegraphics[width = 0.4\textwidth]{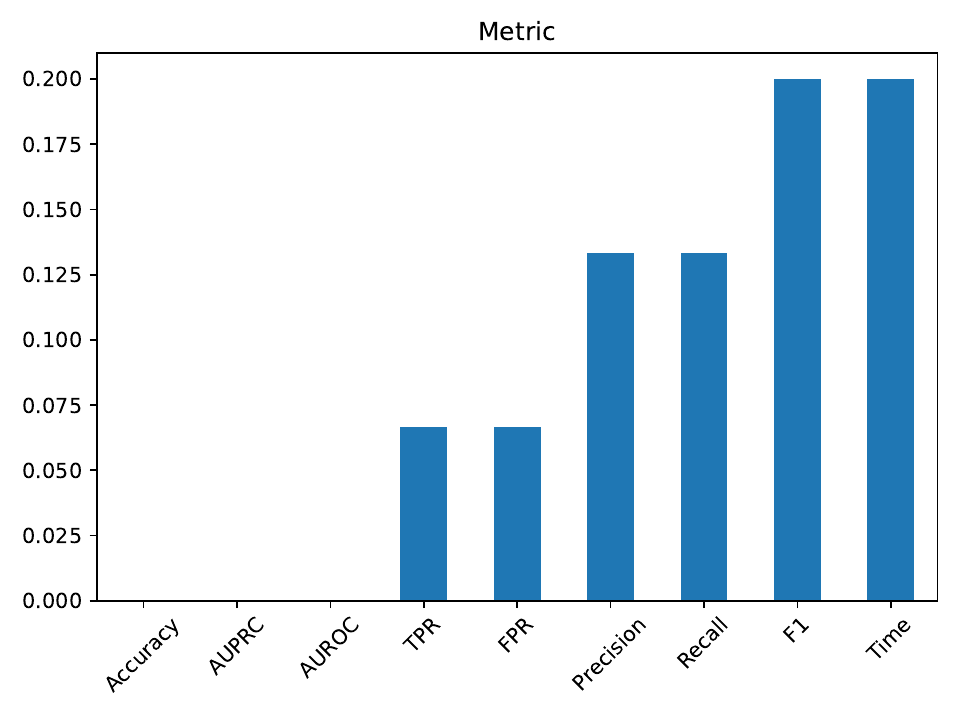}}
\caption{The performance metrics for the top-cited papers.}
\label{fig:topmet}
{}
\end{figure} 

\begin{figure}[h!]
\centering
{\includegraphics[width = 0.4\textwidth]{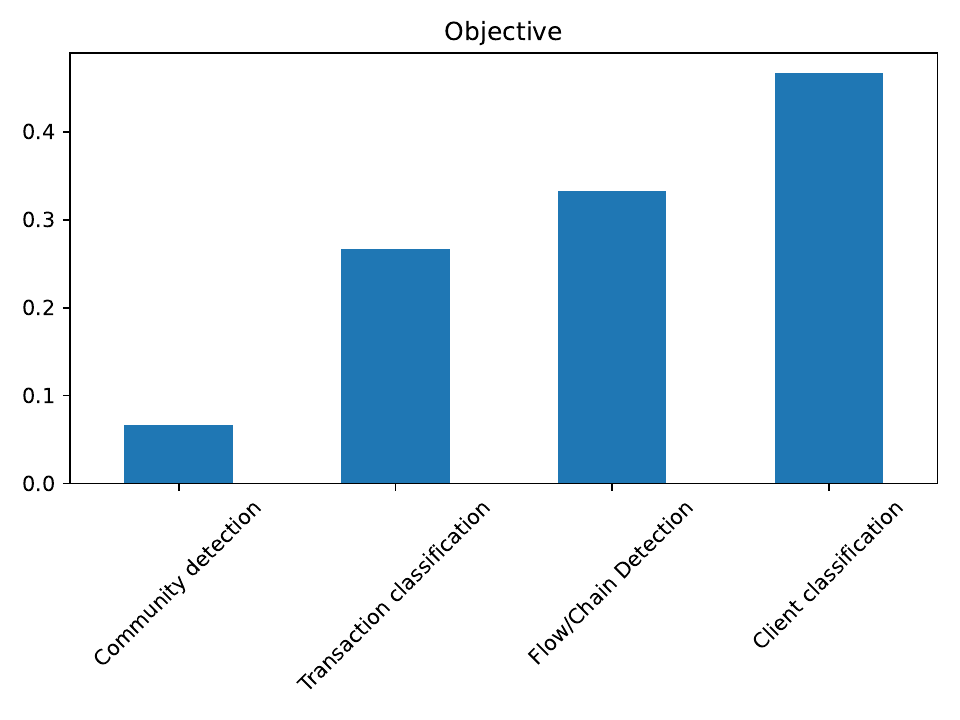}}
\caption{The objective of the top-cited papers.} 
\label{fig:topobj}
{}
\end{figure} 

\begin{figure}[h!]
\centering
{\includegraphics[width = 0.4\textwidth]{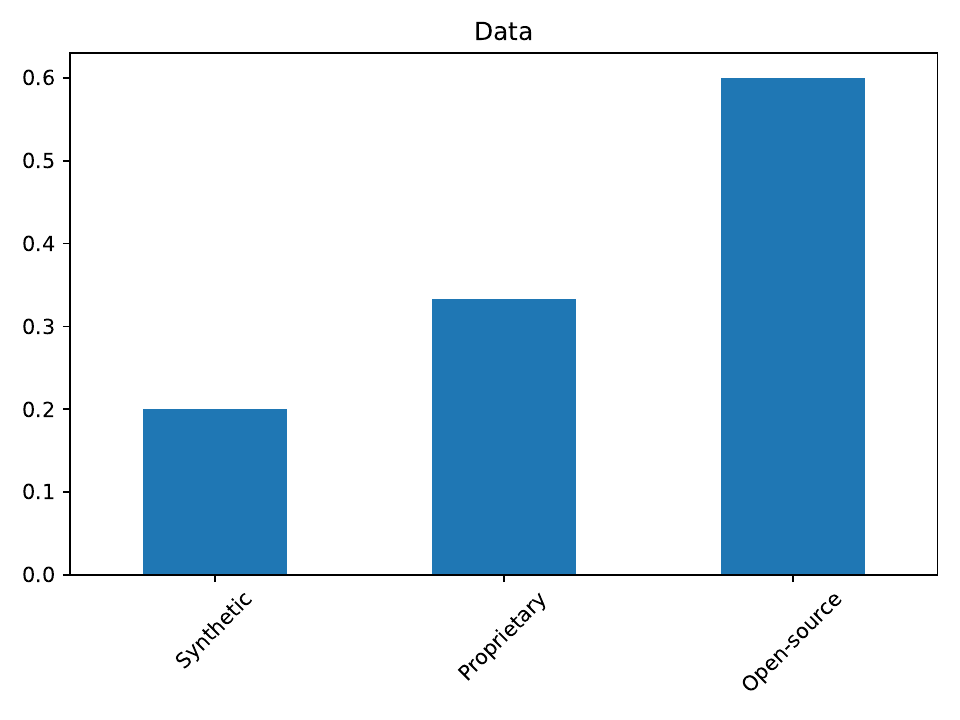}}
\caption{The nature of the data of the top-cited papers.}
\label{fig:topdat}
{}
\end{figure} 
}

\clearpage

\section{Results Threshold-Dependent Metrics}
\label{appendix results percentages}
In this part of the appendix, we give the precision, recall and F1-scores for the models using different thresholds. The thresholds are set such that the observations with the top $0.1\%$, $10\%$ and $p\%$ of scores are classified as money laundering, where $p\%$ is equal to the relative occurrence of the labels. 

\begin{table}
    \centering
    \caption{Threshold-dependent metrics: Precision, recall and F1-score values for the top $0.1\%$ scores over the different methods for the Elliptic data set, based on the test set. The standard deviation is also reported.}
    \label{tab:resTestTD_01_Elliptic}
    {\begin{tabular}{l|ccc}
    \toprule
\bf Methods & \bf Precision & \bf Recall  & \bf F1-score \\ 
\midrule
    Intrinsic features & $0.5396 \pm 0.3093$ & $0.0204 \pm 0.0302$ & $0.0383 \pm 0.0542$ \\
    Egonet features & $0.6833 \pm 0.3187$ & $0.0129 \pm 0.0060$ & $0.0253 \pm 0.0118$ \\
    DeepWalk & $0.5333 \pm 0.3148$ & $0.0101 \pm 0.0059$ & $0.0198 \pm 0.0117$ \\ 
    Node2vec & $0.5167 \pm 0.3352$ & $0.0097 \pm 0.0063$ & $0.0191 \pm 0.0124$ \\
    GCN & $\boldsymbol{1.0000 \pm 0.0000}$ & $0.0189 \pm 0.0007$ & $0.0371 \pm 0.0014$ \\
    GraphSAGE & $0.9624 \pm 0.0615$ & $\boldsymbol{0.0300 \pm 0.0071}$ &  $\boldsymbol{0.0580 \pm 0.0133}$ \\
    GAT & $\boldsymbol{1.0000 \pm 0.0000}$ & $0.0187 \pm 0.0005$ & $0.0367 \pm 0.0010$\\
    GIN & $0.8500 \pm 0.2540$ & $0.0191 \pm 0.0093$ &  $0.0372 \pm 0.0179$\\
    \bottomrule
    \end{tabular}}
    {}
\end{table}

\begin{table}
    \centering
    \caption{Threshold-dependent metrics: Precision, recall and F1-score values for the top $0.1\%$ scores over the different methods for the IBM-AML data set, based on the test set. The standard deviation is also reported.}
    \label{tab:resTestTD_01_IBM}
    {\begin{tabular}{l|ccc}
    \toprule
\bf Methods & \bf Precision & \bf Recall  & \bf F1-score \\ 
\midrule
    Intrinsic features & $0.0050 \pm 0.0087$ & $0.0007 \pm 0.0011$ & $0.0012 \pm 0.0019$ \\
    Egonet features & $0.0079 \pm 0.0085$ & $0.0011 \pm 0.0012$ & $0.0019 \pm 0.0020$ \\
    DeepWalk &  $0.0076 \pm 0.0085$ & $0.0011 \pm 0.0013$ & $0.0019 \pm 0.0022$ \\ 
    Node2vec & $0.0035 \pm 0.0058$ & $0.0006 \pm 0.0009$ & $0.0010 \pm 0.0016$ \\
    GCN &  $\boldsymbol{0.0103 \pm 0.0047}$ & $\boldsymbol{0.8815 \pm 0.3117}$ & $\boldsymbol{0.0203 \pm 0.0092}$\\
    GraphSAGE & $0.0057 \pm 0.0062$ & $0.4919 \pm 0.5188$ & $0.0112 \pm 0.0123$ \\
    GAT &  $0.0054 \pm 0.0087$ & $0.1798 \pm 0.3688$ & $0.0095 \pm 0.0156$ \\
    GIN & $0.0044 \pm 0.0080$ & $0.0102 \pm 0.0131$ & $0.0016 \pm 0.0015$ \\
    \bottomrule
    \end{tabular}}
    {}
\end{table}

\begin{table}
    \centering
   \caption{Threshold-dependent metrics: Precision, recall and F1-score values for the top $10\%$ scores over the different methods for the Elliptic data set, based on the test set. The standard deviation is also reported.}
   \label{tab:resTestTD_10_Elliptic}
    {\begin{tabular}{l|ccc}
    \toprule
\bf Methods & \bf Precision & \bf Recall  & \bf F1-score  \\ 
\midrule
    Intrinsic features & $0.3526 \pm 0.0163$ & $0.6204 \pm 0.0287$ & $0.4497 \pm 0.0208$ \\
    Egonet features & $0.3559 \pm 0.0101$ & $0.6263 \pm 0.0179$ & $0.4539 \pm 0.0129$ \\
    DeepWalk & $0.3517 \pm 0.0050$ & $0.6189 \pm 0.0089$ & $0.4485 \pm 0.0064$ \\ 
    Node2vec & $0.3536 \pm 0.0040$ & $0.6222 \pm 0.0071$ & $0.4509 \pm 0.0051$ \\
    GCN & $0.3471 \pm 0.0272$ & $0.5988 \pm 0.0358$ & $0.4393 \pm 0.0306$\\
    GraphSAGE & $\boldsymbol{0.3675 \pm 0.0118}$ & $\boldsymbol{0.6395 \pm 0.0207}$ & $\boldsymbol{0.4666 \pm 0.0124}$\\
    GAT & $0.3661 \pm 0.0247$ & $0.6388 \pm 0.0288$ & $0.4653 \pm 0.0270$ \\
    GIN & $0.3307 \pm 0.0228$ & $0.5889 \pm 0.0407$ & $0.4234 \pm 0.0274$ \\
    \bottomrule
    \end{tabular}}
    {}
\end{table}

\begin{table}
    \centering
    \caption{Threshold-dependent metrics: Precision, recall and F1-score values for the top $10\%$ scores over the different methods for the IBM-AML data set, based on the test set. The standard deviation is also reported.}
    \label{tab:resTestTD_10_IBM}
    {\begin{tabular}{l|ccc}
    \toprule
\bf Methods & \bf Precision & \bf Recall  & \bf F1-score  \\ 
\midrule
    Intrinsic features & $\boldsymbol{0.0148 \pm 0.0006}$ & $0.1635 \pm 0.0063$ & $\boldsymbol{0.0272 \pm 0.0011}$ \\
    Egonet features & $\boldsymbol{0.0148 \pm 0.0002}$ & $0.1626 \pm 0.0026$ & $0.0271 \pm 0.0004$ \\
    DeepWalk & $0.0145 \pm 0.0003$ & $0.1602 \pm 0.0033$ & $0.0267 \pm 0.0005$ \\ 
    Node2vec & $\boldsymbol{0.0148 \pm 0.0006}$ & $0.1630 \pm 0.0065$ & $0.0271 \pm 0.0011$ \\
    GCN & $0.0109 \pm 0.0030$ & $0.9814 \pm 0.0374$ & $0.0214 \pm 0.0058$ \\
    GraphSAGE & $0.0121 \pm 0.0026$ & $0.9761 \pm 0.0257$ & $0.0238 \pm 0.0051$ \\
    GAT & $0.0054 \pm 0.0074$ & $\boldsymbol{0.1993 \pm 0.3722}$ & $0.0104 \pm 0.0144$\\
    GIN & $0.0081 \pm 0.0077$ & $0.0947 \pm 0.0793$ & $0.0149 \pm 0.0141$ \\
    \bottomrule
    \end{tabular}}
    {}
\end{table}

\begin{table}
    \centering
    \caption{Threshold-dependent metrics: Precision, recall and F1-score values for the top $p=2\%$ scores over the different methods for the Elliptic data set, based on the test set. The standard deviation is also reported.} 
    \label{tab:resTestTD_p_Elliptic}
    {\begin{tabular}{l|ccc}
    \toprule
\bf Methods & \bf Precision & \bf Recall  & \bf F1-score  \\ 
\midrule
    Intrinsic features & $0.3244 \pm 0.0143$ & $0.6303 \pm 0.0277$ & $0.4283 \pm 0.0188$ \\
    Egonet features & $0.3271 \pm 0.0098$ & $0.6357 \pm 0.0190$ & $0.4319 \pm 0.0129$ \\
    DeepWalk & $0.3220 \pm 0.0052$ & $0.6258 \pm 0.0102$ & $0.4252 \pm 0.0069$ \\ 
    Node2vec & $0.3241 \pm 0.0046$ & $0.6299 \pm 0.0089$ & $0.4280 \pm 0.0060$ \\
    GCN & $0.3209 \pm 0.0284$ & $0.6230 \pm 0.0371$ & $0.4234 \pm 0.0325$\\
    GraphSAGE &$0.3343 \pm 0.0130$ & $0.6558 \pm 0.0166$ & $0.4427 \pm 0.0121$ \\
    GAT & $\boldsymbol{0.3419 \pm 0.0084}$ &  $\boldsymbol{0.6613 \pm 0.0212}$ &  $\boldsymbol{0.4507 \pm 0.0109}$ \\
    GIN & $0.3034 \pm 0.0169$ & $0.5949 \pm 0.0285$ & $0.4018 \pm 0.0205$ \\
    \bottomrule
    \end{tabular}}
    {}
\end{table}

\begin{table}
    \centering
    \caption{Threshold-dependent metrics: Precision, recall and F1-score values for the top $p=0.11\%$ scores over the different methods for the Elliptic data set, based on the test set. The standard deviation is also reported.}
    \label{tab:resTestTD_p_ibm}
    {\begin{tabular}{l|ccc}
    \toprule
\bf Methods & \bf Precision & \bf Recall  & \bf F1-score  \\ 
\midrule
    Intrinsic features & $0.0056 \pm 0.0090$ & $0.0008 \pm 0.0012$ & $0.0014 \pm 0.0021$ \\
    Egonet features & $0.0085 \pm 0.0077$ & $0.0012 \pm 0.0011$ & $0.0021 \pm 0.0019$ \\
    DeepWalk & $0.0080 \pm 0.0076$ & $0.0013 \pm 0.0013$ & $0.0023 \pm 0.0021$ \\ 
    Node2vec & $0.0047 \pm 0.0063$ & $0.0008 \pm 0.0010$ & $0.0013 \pm 0.0018$ \\
    GCN & $\boldsymbol{0.0103 \pm 0.0046}$  & $\boldsymbol{0.8811 \pm 0.3119}$ & $\boldsymbol{0.0203 \pm 0.0089}$ \\
    GraphSAGE & $0.0055 \pm 0.0061$ & $0.4907 \pm 0.5175$ & $0.0108 \pm 0.0121$ \\
    GAT & $0.0047 \pm 0.0078$ & $0.1763 \pm 0.3656$ & $0.0085 \pm 0.0146$ \\
    GIN & $0.0058 \pm 0.0084$ & $0.0091 \pm 0.0116$ & $0.0019 \pm 0.0016$ \\
    \bottomrule
    \end{tabular}}
    {}
\end{table}

\end{document}